\documentclass[a4paper,11pt]{article}
\pdfoutput=1 % if your are submitting a pdflatex (i.e. if you have
\usepackage{jheppub} % for details on the use of the package, please
\usepackage[T1]{fontenc} % if needed
\usepackage[dvipsnames]{xcolor} %Must come before tikz!
\colorlet{cite}{LimeGreen!50!Green}
\usepackage[utf8]{inputenc}
\usepackage{soul}
\usepackage{tikz}  % for diagrams
\usetikzlibrary{arrows,positioning}
\tikzset{ 
  baseline=-2.3pt,
  text height=1.5ex, text depth=0.25ex,
  >=stealth,
  node distance=2cm,
  mid/.style={fill=white,inner sep=2.5pt},
}
\usepackage{lmodern}
\usepackage{tikz-cd}
\usepackage{amsthm, amssymb, amsfonts}
\usepackage{multirow}
\usepackage{cancel}
\usepackage{diagbox}
\usepackage{makecell}
\input xypic

\usepackage{graphicx,caption,subcaption}
\usepackage{braket}  % For nice sets
\usepackage{cleveref}

\providecommand{\abs}[1]{\lvert#1\rvert}

\DeclareMathOperator{\Tr}{Tr}

\newcommand{\old}[1]{}

\theoremstyle{plain}	% 'plain' is the default.  The others are 'definition' and 'remark'.
\newtheorem{theorem}{Theorem} % putting [section] on the end here tells latex to number the theorem environment within sections, ie. Theorem 2.3 for the third theorem in section 2.
\newtheorem{lemma}[theorem]{Lemma} 

\newtheorem*{theorem*}{Theorem}

\theoremstyle{definition}

\newtheorem*{conjecture*}{Conjecture}
\theoremstyle{remark}

\newtheorem{definition}[theorem]{Definition}

\newtheorem*{lemma*}{Lemma}

\usepackage{amsmath,amssymb, xcolor}
\title{\boldmath Torsion in M2-brane theory}

 \author[a,b]{M.P. García del Moral,}
 \author[c]{C. las Heras,}
 \author[d]{A. Restuccia,}\note{These authors contributed equally to this work.} 
\affiliation[a]{Departamento de Química, Área de Física, Universidad de la Rioja,\\
 C/ Madre de Dios 53, Logroño 26006, España.}
\affiliation[b]{Instituto de Investigación en Computación Científica (SCRIUR), Universidad de la Rioja,\\ C/ Madre de Dios 53, Logroño 26006, España.}
\affiliation[c]{Instituto de Física Teórica UAM/CSIC, Universidad Autónoma de Madrid, \\
 C/ Nicolás Cabrera 13-15, Cantoblanco, Madrid 28040, Spain}
 \affiliation[d]{Departamento de Física, Universidad de Antofagasta,\\ Universidad de Antofagasta, Campus Coloso, Aptdo 02800, Antofagasta, Chile}
\emailAdd{m-pilar.garciam@unirioja.es}
\emailAdd{camilo.lasheras@ift.csic.es}
\emailAdd{alvaro.restuccia@uantof.cl}

\abstract{We determine the role of torsion in the local and global geometrical description of M2-branes with fluxes and parabolic monodromies. The monodromy corresponds to a representation of the fundamental group of the base manifold into the parabolic subgroup of $\mbox{SL}(2,\mathbb{Z})$, the group of isotopy classes of area preserving diffeomorphisms. These are supersymmetric M2-branes with a quantum discrete spectrum with finite multiplicity. The global description of these QM2-branes is given by twisted torus bundles with monodromy. They are classified by $\mbox{H}^2(\Sigma,\mathbb{Z}_\rho)=\mathbb{Z}\oplus \mathbb{Z}_k$, or equivalently, by the coinvariants associated with the parabolic monodromy subgroup. We generalize previous constructions in two different ways. The first one considers parabolic monodromies with $k>1$. This will allow us to identify torsion cycles of order greater than one. We find that there are well-defined nilmanifolds in three, four, and five dimensions contained in the global description of these M2-branes. These nilmanifolds are in correspondence with the nilpotent Lie algebras $g_{3,1}$, $g_{3,1}\oplus g_1$, and $g_{3,1}\oplus g_2$. All these nilmanifolds have torsion cycles of order $k$ contained in the compact sector of the target-space of the M2-brane. The torsion is also manifest in the equivalence classes of M2-brane twisted torus bundles, namely coinvariants. The second one is that we analyze how the torsion acts on the coinvariants of the base manifold. Together with the flux condition, the torsion defines explicitly the number of coinvariants for a given flux and monodromy $k$.}

\begin{document}
\emergencystretch 3em
\hypersetup{pageanchor=false}
\makeatletter
\let\old@fpheader\@fpheader
\preprint{IFT-UAM/CSIC-25-34}

\makeatother
 \maketitle
\flushbottom

\section{Introduction}
\label{sec:intro}

In this work, we analyze the role of torsion in the local and global description of QM2-branes on $M_9\times T^2$ with parabolic monodromies, where $M_9$ corresponds to the Minkowski space-time in nine dimensions. We identify several nilmanifolds in low dimensions, which are contained in the M2-brane bundle description, implying torsion cycles in the compactification manifold. We discuss the contribution of these torsion cycles in the mass operator and the equivalence classes of M2-branes twisted torus bundles.

Solvmanifolds have been widely used in string compactifications to obtain Minkowski, de Sitter, or Anti-de Sitter vacua \cite{Silverstein, Andriot1, Andriot2, Grana, Grana2}. When solvmanifolds are associated with nilpotent Lie groups, they are known as nilmanifolds. Low-dimensional nilmanifolds are well-known and completely characterized. They correspond with the non-isomorphic nilpotent Lie algebras in low dimensions. Besides the abelian cases associated with $T^3$, $T^4$, and $T^5$, there are: one three-dimensional nilmanifold, two four-dimensional nilmanifolds, and eight five-dimensional nilmanifolds \cite{Bock}. These ones have a nontrivial torsion in their homology group.

The three-dimensional nilmanifold is known as the Heisenberg nilmanifold and is associated with the Heisenberg Lie algebra in three dimensions. This nilmanifold has two consistent geometric descriptions. It can be seen as a $\mbox{U}(1)$ principal bundle over a torus with $c_1=k$, or by its Mostow fibration, which is a torus bundle over a circle with parabolic monodromy, where $k$ now characterizes the off-diagonal element of the $2\times 2$ monodromy matrix \cite{Bock,Aschieri}. The four-dimensional nilmanifolds correspond to: Primary Kodaira Surfaces, associated with the decomposable nilpotent Lie algebra in four dimensions; and the extended Heisenberg nilmanifold (or filiform nilmanifold), associated with the indecomposable nilpotent Lie algebra in four dimensions. These nilmanifolds also have two consistent geometric descriptions, given by symplectic torus bundles over a torus or as circle bundles over the Heisenberg nilmanifold \cite{Bock, Latorre,Fernandez,Hull11,Palais,Belegradek}. In the particular case where $k=1$, the Primary Kodaira Surface corresponds to the Kodaira--Thurston manifold, which is the first known example of a symplectic manifold, not Kahler \cite{Kodaira,Latorre,Kahn,Hull10}. Five-dimensional nilmanifolds are in correspondence with the two decomposable nilpotent Lie algebras and with the 6 indecomposable nilpotent Lie algebras in five dimensions. Among these nilmanifolds, we will be particularly interested in those whose Lie algebra is decomposable and associated with the Heisenberg Lie algebra in three dimensions \cite{Bock}.

Type II supergravities with 32 supercharges in nine dimensions were characterized in \cite{Bergshoeff5}. The UV description associated with these supergravities has been considered in terms of generalized Scherk--Schwarz reduction of type IIB superstring, equivalently, F-theory on three-dimensional solvmanifolds \cite{Hull8,Hull4}, or in terms of M2-branes with fluxes and monodromy on $M_9\times T^2$ \cite{Hull8,Hull4,mpgm2,mpgm7,mpgm23,mpgm24}. We denote them by QM2-branes because they have a purely discrete supersymmetric spectrum with finite multiplicity. In the original work, the M2-brane was considered on $M_9\times T^2$ with a topological restriction on the embedding maps of the theory \cite{Restuccia}. It implies the existence of a nontrivial flux condition on the worldvolume of the theory. The discreteness of the supersymmetric spectrum of the $\mbox{SU}(N)$ regularized model was shown in \cite{Boulton}. Consequently, it may describe microscopical degrees of freedom of, at least, a sector of M-theory. This sector was shown to be equivalent to the M2-brane on $M_9\times T^2$ with a nontrivial flux condition on $T^2$ \cite{mpgm6}. Other backgrounds of M2-branes with a good quantum description are given by \cite{mpgm26,mpgm11,mpgm29,mpgm30}. 

QM2-branes on $M_9\times T^2$ are geometrically described by twisted torus bundles. Given a torus bundle with base manifold $M$, fiber $F$, and structure group $G$, where $F$ is a torus $T^2$ and $G$, the symplectomorphisms acting on $F$, we may also consider the principal bundle $(P,M,G)$ and the associated bundle $E$. In \cite{mpgm10}, we introduced a one-form symplectic connection and a $\mbox{U}(1)$ one-form connection on $M$, the pullback of a connection on $E$. The $\mbox{U}(1)$ structure defines a twisted torus. It is in this sense that we refer to twisted torus bundle. The monodromy construction is in terms of a homomorphism of the fundamental group of M onto the group of isotopy classes of symplectomorphisms on the fiber T2. In our work, $M$ is also a $T^2$ torus. From \cite{mpgm24}, we know that the transformation within equivalent M2-brane bundles reproduces the gauge symmetry group, while the transformation between inequivalent bundles predicts the U-duality group as a subgroup of $\mbox{SL}(2,\mathbb{Z})$. In this work, we relate the coinvariant structure of the target to the coinvariant structure of the base manifold, which depends on the flux condition. Moreover, for the parabolic subgroup of $\mbox{SL}(2,\mathbb{Z})$, while in \cite{mpgm23,mpgm24} it was only considered the case where $k=1$, with $k$ the off-diagonal element in the monodromy matrix, we analyze here the M2-brane for all $k>1$ and determine all the torsion cycles of order greater than one. Some works considering torsion in (co)homology in different contexts are \cite{Marchesano2,Marchesano3,Marchesano4,Weigand} and references therein.

This work is organized as follows: in section \ref{Sec1}, we review basic notions of nilmanifolds in three, four, and five dimensions. We make special emphasis on those nilmanifolds, whose correspondent nilpotent Lie algebra is decomposable in terms of the Heisenberg Lie algebra in three dimensions. In section \ref{Sec2}, the worldvolume and bundle description of QM2-branes with monodromy is given, as well as their relation with type IIB gauged supergravity in nine dimensions. New results are presented in sections \ref{Sec3}, \ref{Sec4}, and \ref{Sec5}. In section \ref{Sec3}, we identify different nilmanifolds in low dimensions that are contained in the bundle description of the QM2-branes with parabolic monodromies. In section \ref{Sec4}, we analyze the role that torsion plays in the Hamiltonian of the QM2-brane with parabolic monodromy and in the equivalence classes of M2-brane twisted torus bundles. We generalize previous works by considering the role of the torsion of order greater than one, as well as the role of the coinvariant of the base, associated with the flux condition. Finally, we discuss our results in section \ref{Sec5}.

\section{Some relevant features of nilmanifolds in low dimensions}
\label{Sec1}

This section will introduce some basic notions on orientable solvmanifolds in low dimensions. In this work, we will place special emphasis on nilmanifolds in three, four, and five dimensions. Except for the abelian cases, all these nilmanifolds have torsion in the homology. Solvmanifolds in general have been extensively used in the context of string compactifications to obtain Minkowski, anti-de Sitter, and de Sitter vacua \cite{Silverstein, Andriot1, Andriot2, Grana, Grana2}. For a more detailed analysis of the properties of low-dimensional solvmanifolds, see \cite{Bock}.

\subsection{Solvmanifolds in a nutshell.}

    A \textit{solvmanifold} is a compact homogeneous space $$\Lambda \backslash G,$$ where $G$ is a (connected and) simply connected solvable Lie group and $\Lambda$ a lattice in $G$, i.e., a discrete co-compact subgroup. Solvmanifolds are, in general, fibrations over tori.
    
    When $G$ is nilpotent, then $\Lambda \backslash G$ is a \textit{nilmanifold}. Every nilmanifold is a solvmanifold, but some solvmanifolds are not diffeomorphic to nilmanifolds. 
    
    While the existence of a lattice is guaranteed for a nilpotent Lie group, there is no such criteria for general solvmanifolds. But this issue can be avoided if we consider almost Abelian solvmanifolds. In that case, we have that $G$ have the structure of a semi-direct product (See Appendix \ref{Ape1})
$$G=\mathbb{R}^{d-1}\rtimes_\varphi \mathbb{R}$$
with $\varphi$ a continuous one-parameter left group action $\varphi:\mathbb{R}\rightarrow \mbox{Aut}(\mathbb{R}^{d-1})$. We can regard $\varphi$ as a matrix $\varphi_x\in \mbox{GL}(d-1,\mathbb{R})$ for each $x\in\mathbb{R}$. Then, $G$ admits a lattice $\Lambda_G$ if and only if there exists $x_0\in\mathbb{R}^\times$ such that 
$$\Sigma^{-1}\varphi_{x_0}\Sigma = M$$
with $M\in \mbox{SL}(d-1,\mathbb{Z})$ and $\Sigma\in \mbox{GL}(d-1,\mathbb{R})$. Indeed, this condition also ensures the compactness of the solvmanifold \cite{Bock,Aschieri}.

The corresponding Lie algebra associated with a solvmanifold is defined by
    \begin{eqnarray}
      \left[ E_b,E_c\right]={f^a}_{bc}E_a ,
    \end{eqnarray}
where $\left\lbrace E_a \right\rbrace$ is a vector basis and ${f^a}_{bc}$ are the structure constants. There is a set of dual one-forms $\left\lbrace \eta^a\right\rbrace$ satisfying a Maurer--Cartan equation
    \begin{eqnarray}
      d\eta^a=-\frac{1}{2}{f^a}_{bc}\eta^b\wedge \eta^c. 
    \end{eqnarray}
    These one-forms are globally well-defined on the manifold. Its topological structure is given by the Maurer--Cartan equation. 

A large class of twisted tori of interest in string compactifications comes in the form of fibrations over tori. These are solvmanifolds. In this work, we will be especially interested in twisted tori which are almost Abelian solvmanifolds.
\begin{definition}[twisted tori]
    Let $G$ be a locally compact group that admits a cocompact discrete subgroup, $\Lambda_G$, i.e., a lattice in $G$, acting on $G$ by left multiplication. The quotient space
    $$\mathbb{T}_{\Lambda_G}:= \Lambda_G \backslash G$$
    is a \textit{twisted torus}.
\end{definition}
 Some examples of twisted torus are: the $d$-dimensional torus $\mathbb{T}_{\mathbb{Z}^d}=\mathbb{Z}^d\backslash \mathbb{R}^d$, nilmanifolds, orbifolds as the three-dimensional ADE orbifolds $\mathbb{T}_\Gamma=\Gamma\backslash \mathbb{S}^3$ for $\Gamma\subset SU(2)$, or the Lens spaces $\mathbb{T}_{\mathbb{Z}_n}=\mathbb{Z}_n\backslash \mathbb{S}^3$ 

The fibrations underlying the twisted tori, as almost Abelian solvmanifolds, are called Mostow bundles \cite{Mostow}.

    \begin{theorem}[Mostow bundles]
Let $\Lambda_G$ be a lattice in a connected and simply-connected solvable Lie group $G$ and $\mathbb{T}_{\Lambda_G}=\Lambda_G\backslash G$ the associated solvmanifold. Let $N$ be the nilradical of $G$. Then $\Lambda_G N$ is a closed subgroup of $G$, $\Lambda_N:=\Lambda_G\cap N$ is a lattice in $N$, and $\Lambda_G N\backslash G$ is a torus. It follows that the twisted torus $\mathbb{T}_{\Lambda_G}$ is a fibration over this torus with a nilmanifold fiber given by:
$$\Lambda_N \backslash N=\Lambda_G\backslash\Lambda_G N \longrightarrow \mathbb{T}_{\Lambda_G}\longrightarrow \Lambda_G N\backslash G.$$
\end{theorem}
That is, a nilmanifold fibered over a torus. The structure group of the Mostow bundle is $\Lambda_{G_0}\backslash \Lambda_G N$ where $\Lambda_{G_0}$ is the largest subgroup of $\Lambda_G$ which is normal in $\Lambda_G N$.

The corresponding Mostow bundle, associated with an almost Abelian solvmanifold is
$$\Lambda_N\backslash N \backsimeq\mathbb{T}^{d-1}\longrightarrow \mathbb{T}_{\Lambda_G}\longrightarrow \Lambda_G N \backslash G \backsimeq \mathbb{T}$$
which corresponds with a torus bundle over a circle with monodromy $M$ in the mapping class group $\mbox{SL}(d-1,\mathbb{Z})$ of orientation-preserving automorphisms up to the homotopy of the torus fibers.

Let us note that if $x$ is a coordinate over the circle and $(y^1,\dots,y^d)$ coordinates over the torus, we have that
\begin{eqnarray}
    \left[E_a,E_x\right]={M_a}^bE_b \,, \mbox{  and  } \, \left[E_a,E_b\right] = 0
\end{eqnarray}
with $E_1,\dots,E_d,E_x$ the generators of the Lie algebra and $M$ is the mass matrix such that $\varphi(x)=\exp{(Mx)}$. The corresponding basis of left-invariant one-forms are
\begin{eqnarray}
    \eta^x&=&dx, \\
    \eta^a &=& {\varphi(x)^a}_bdy^b
\end{eqnarray}
    such that
    \begin{eqnarray}
        d\eta^x &=& 0 , \\
        d\eta^a &=& -{M^a}_b\eta^x\wedge\eta^b
    \end{eqnarray}

Now, let us consider some particular nilmanifolds in three, four, and five dimensions that will be relevant in our work.

    \subsection{Nilmanifolds in low dimensions.}
In the following, we review relevant features for our work of nilmanifolds formulated in three, four, and five dimensions. In the global description of the QM2-brane in $M_9\times T^2$ (see section \ref{Sec2}), we consider a fiber bundle with the target-space $M_9\times T^2$ fibered over a base manifold describing the worldvolume of the M2-brane. The topologically nontrivial part corresponds to a $T^2$ fibered over the Riemann surface of genus one, associated with the spatial directions of the worldvolume. Since the M2-brane contains fluxes, there is an extra $\mbox{U}(1)$ that geometrically corresponds to an extra $S^1$. For this reason, we are interested in describing all the geometrical structures contained in the nontrivial five-dimensional M2-brane bundle. 
    \subsubsection{Three dimensions}
Following theorem (5.3) from \cite{Bock}, we observe that three-dimensional solvmanifolds $G/\Gamma$ are contained in table \ref{tab1era}. 
\begin{table}[]
    \centering
    \begin{tabular}{|c|c|c|c|c|}
 \hline
     & $b_1(G/\Gamma)$ & $G$ & Nilmanifold & Completely solvable  \\
    \hline
    a) & 3 & $\mathbb{R}^3$ &yes & yes \\
       \hline
    b) & 2 & $\mathcal{H}_3$ &yes & yes\\
       \hline
    c) & 1 & $\mbox{ISO}(1,1)$ &no & yes \\
       \hline
    d) & 1 & $\mbox{ISO}(2)$ &no & no \\
    \hline
    \end{tabular}
    \caption{Solvmanifolds in three dimensions from \cite{Bock}, where $\mathcal{H}_3$, $\mbox{ISO}(1,1)$ and $\mbox{ISO}(2)$ corresponds to the Heisenberg, Poincare, and Euclidean groups in three dimensions.}
    \label{tab1era}
\end{table}
These solvmanifolds are in correspondence with the isomorphism classes of Lie algebras of connected solvable Lie groups that possess lattices (see table \ref{tab2da}). 
\begin{table}[]
    \centering
    \begin{tabular}{|c|c|c|c|}
 \hline
     & Algebra & $\left[ E_i,E_j\right]$ & Completely solvable  \\
    \hline
    a) & $3g_1$ &  & abelian \\
       \hline
    b) & $g_{3,1}$ & $\left[E_2,E_3\right]=E_1 $ & nilpotent\\
       \hline
    c) & $g_{3,4}^{-1}$ & $\left[E_1,E_3\right]=E_1 $  , $\left[E_2,E_3\right]=-E_2 $ & yes \\
       \hline
    d) & $g_{3,5}^0$ & $\left[E_1,E_3\right]=-E_2 $  , $\left[E_2,E_3\right]=E_1 $  & no \\
    \hline
    \end{tabular}
    \caption{Non-isomorphic algebras in three dimensions from \cite{Bock}}
    \label{tab2da}
\end{table}

    All solvmanifolds in three dimensions are almost abelian solvmanifolds. Indeed, we have that $$G=\mathbb{R}^2\rtimes_\varphi\mathbb{R}.$$
     
    For $\varphi_x=\mathbb{I}_2 \, \forall \, x\in\mathbb{R}$, then $G$ is abelian and the solvmanifold is a 3-torus. The remaining cases correspond to
$$\varphi:\mathbb{R}\longrightarrow \mbox{SL}(2,\mathbb{R}).$$  Consequently, they are classified into conjugacy classes given by the trace.

The Mostow fibrations of these solvmanifolds are given by torus bundles over a circle with monodromy in $\mbox{SL}(2,\mathbb{Z})$. That is
$$T^2\longrightarrow G/\Gamma \longrightarrow S^1$$
with $$\rho:\Pi_1(S^1)\longrightarrow \mbox{SL}(2,\mathbb{Z}).$$
    \begin{table}
        \centering
        \begin{tabular}{|c|c|c|c|}
        \hline
             &  Monodromy & $\mbox{H}_1(\cdot,\mathbb{Z})$ & Completely solvable \\
             \hline
           a) & $\begin{pmatrix}
               1 & 0 \\
               0 & 1
           \end{pmatrix}$ & $\mbox{H}_1(T^3,\mathbb{Z})=\mathbb{Z}\oplus\mathbb{Z}\oplus\mathbb{Z}$ & Abelian \\
           \hline
           b) & $\begin{pmatrix}
               1 & k \\
               0 & 1
           \end{pmatrix}$ & $\mbox{H}_1(T_p^3,\mathbb{Z})=\mathbb{Z}\oplus\mathbb{Z}\oplus\mathbb{Z}_k$ & Nil \\
           \hline
           c) & $\begin{pmatrix}
               n & -1 \\
               1 & 0
           \end{pmatrix}$ & $\mbox{H}_1(T_h^3,\mathbb{Z})=\mathbb{Z}\oplus\mathbb{Z}_{n-2}$ & Solv \\
           \hline
           d) & $\begin{pmatrix}
               -1 & 0 \\
               0 & -1
           \end{pmatrix}$ & $\mbox{H}_1(T_{\mathbb{Z}_2}^3,\mathbb{Z})=\mathbb{Z}\oplus\mathbb{Z}_2\oplus\mathbb{Z}_2$ & Solv \\
           \hline
            d) & $\begin{pmatrix}
               0 & -1 \\
               1 & 0
           \end{pmatrix}$ & $\mbox{H}_1(T_{\mathbb{Z}_4}^3,\mathbb{Z})=\mathbb{Z}\oplus\mathbb{Z}_2$ & Solv \\
           \hline
         d) & $\begin{pmatrix}
               0 & -1 \\
               1 & -1
           \end{pmatrix}$ & $\mbox{H}_1(T_{\mathbb{Z}_6}^3,\mathbb{Z})=\mathbb{Z}\oplus\mathbb{Z}_3$ & Solv \\
           \hline
        \end{tabular}
        \caption{Non-isomorphic twisted tori in three dimensions}
        \label{tab3era}
    \end{table}
Therefore, the $T^3$ correspond to trivial monodromy (case (a) in tables \ref{tab1era}, \ref{tab2da} and \ref{tab3era})), the Heisenberg nilmanifold (case (b)) corresponds to parabolic monodromies ($\vert \Tr(\rho)\vert=2$), and the twisted tori with hyperbolic ($\vert Tr(\rho)\vert>2$), and elliptic ($\vert Tr(\rho)\vert<2$) monodromies are associated with cases c) and d), respectively. The last two cases are also denoted in the literature by $\varepsilon_{1,1}$ and $\varepsilon_2$, respectively. We will not consider them any further in this work.

 The first homology group of these solvmanifolds is given by
$$\mbox{H}_1(\mathbb{T}_{\Lambda_G},\mathbb{Z})\simeq \mathbb{Z}\oplus\mathbb{Z}^{2-r}\oplus \bigoplus_{i=1}^r\mathbb{Z}_{m_i}$$
where $r$ is the rank of the matrix $A:=M-\mathbb{I}_{2}$ and $$m_i=\frac{d_i(A)}{d_{i-1}(A)}$$ with $d_i(A)$ the greatest common divisor of all $i\times i$ minors of $A$, with $d_0(A)=1$.

All these solvmanifolds are technically known as twisted tori. However, most authors use the name of twisted torus for the Heisenberg nilmanifold. Solvmanifolds are widely used in the context of string compactification; see, for example, \cite{Silverstein,Andriot1,Andriot2,Grana,Grana2}.

It can be seen from Table (\ref{tab3era}) that nontrivial monodromies imply the existence of torsion cycles in the homology. We will have one torsion cycle for the Heisenberg nilmanifold and two torsion cycles in all other cases. In this work, we will focus on the Heisenberg nilmanifold in three dimensions. Despite having a geometrical description of a torus bundle over a circle with parabolic monodromy, it has the nice property that it can also be described as a nontrivial $\mbox{U}(1)$ principal bundle over a torus $$\mbox{U}(1)\to E\to T^2, \, \mbox{  with  } c_1=k.$$ In this case, $k$ from case (b) in table \ref{tab3era}, corresponds to the first Chern number of the principal bundle.

\paragraph{Heisenberg nilmanifold:} 
The Heisenberg group in three dimensions is the unique three-dimensional connected and simply-connected non-abelian nilpotent Lie group.
$$\mathcal{H}_3 := \left\lbrace\begin{pmatrix}
    1 & x & y \\
    0 & 1 & z \\
    0 & 0 & 1
\end{pmatrix}\quad \vert \quad x,y,z\in\mathbb{R} \right\rbrace $$
This group is diffeomorphic to $\mathbb{R}^3$ with the following group multiplication,
$$(x,y,z)(x',y',z')=(x+x',y+y'+xz',z+z')$$
and the inverse element is given by
$$(x,y,x)^{-1}=(-x,xz-y,-z).$$

The Mostow bundle structure is based on the semi-direct product
\begin{eqnarray}
    \mathcal{H}_3 = \mathbb{R}^2 \rtimes_\varphi \mathbb{R}
\end{eqnarray}
with $\varphi_x=\begin{pmatrix}
    1 & x \\ 0 & 1
\end{pmatrix}$. In this case, it is evident that when $x_0=k\in\mathbb{Z}$, then the matrix $\varphi_{x_0}$ is integer-valued. 

The Heisenberg nilmanifold is the compact space obtained by the quotient of the Heisenberg group with respect to the lattice given by the discrete Heisenberg group  $$\mathcal{H}_3/\mbox{Heis}_k(3;\mathbb{Z}).$$ 
It can be seen that every lattice of the Heisenberg group in three dimensions is isomorphic to \cite{Bock}
$$\mbox{Heis}_k(3;\mathbb{Z})\equiv\Gamma_{3,k}:=\Gamma_{3,k}(\mathbb{Z}):=\left\lbrace \begin{pmatrix}
    1 & x & \frac{y}{k} \\
    0 & 1 & z \\
    0 & 0 & 1
\end{pmatrix}\quad \vert \quad x,y,z\in\mathbb{Z} \, \mbox{ and } \, k\in\mathbb{N}_+\right\rbrace.$$

It will be useful to denote the center and the automorphisms groups of $\mathcal{H}_3$  as \cite{Bock}
\begin{eqnarray}
    Z(\mathcal{H}_3) &=& \left\lbrace (0,0,z) \quad \vert \quad z\in\mathbb{R} \right\rbrace , \\
    \mbox{Aut}(\mathcal{H}_3) &=& \mbox{GL}(2,\mathbb{R})\times \mathbb{R}^2.
\end{eqnarray}
The integer $k\in\mathbb{Z}$ corresponds to the non-diagonal element of the monodromy matrix from the point of view of the Mostow bundle, and it corresponds to the first Chern number $c_1=k$ from the nontrivial $\mbox{U}(1)$ principal bundle description.

The equivalence relation is given by the left action of $\mbox{Heis}_k(3;\mathbb{Z})$ which leads to the following coordinate identification
\begin{eqnarray}
    (x,y,z)&\longrightarrow& (x+k,y+kz,z) \\
    (x,y,z)&\longrightarrow& (x,y+1,z) \\
    (x,y,z)&\longrightarrow& (x,y,z+1)
\end{eqnarray}
The set of one-forms that are globally well defined on the Heisenberg nilmanifold are given by
\begin{eqnarray}
    \eta^1 &=& dx, \\
    \eta^2 &=& dz, \\
    \eta^3 &=& dy + kxdz
\end{eqnarray}
such that Maurer--Cartan equation is satisfied $d\eta^3 = k\eta^1\wedge \eta^2$. In this notation, $x\in S^1$ while $(y,z)\in T^2$. 

The Heisenberg nilmanifold is characterized by 
$$\mbox{H}_1(T_p^3,\mathbb{Z})=\mathbb{Z}\oplus \mathbb{Z} \oplus \mathbb{Z}_k=\mbox{H}^2(T_p^3,\mathbb{Z})$$
and it is related by T-duality to a $T^3$ with $H$-flux and T-folds.

\subsubsection{Four dimensions}

Nonisomorphic solvmanifolds in four dimensions are known. Following theorem (6.2) from \cite{Bock}, we have that there are three nilmanifolds in four dimensions, given by table \ref{tab4ta}. The Lie algebras that are in correspondence with these nilmanifolds are given in table \ref{tab5ta}.
\begin{table}[]
    \centering
    \begin{tabular}{|c|c|c|c|c|}
 \hline
     & $b_1(G/\Gamma)$ & Symplectic & Complex & Kahler  \\
    \hline
    a) & 4 & yes & Torus & yes \\
       \hline
    b) & 3 & yes & PKS & no \\
       \hline
    d) & 2 & yes & no & no \\
    \hline
    \end{tabular}
    \caption{Nilmanifolds in four dimensions from \cite{Bock}. PKS stands for Primary Kodaira Surfaces.}
    \label{tab4ta}
\end{table}
Nilmanifolds in four dimensions with $b_1=4$ correspond to the 4-torus, nilmanifolds with $b_1=3$ correspond to Primary Kodaira surfaces, and nilmanifolds with $b_1=2$ are known in the literature as extended Heisenberg nilmanifolds or filiform nilmanifolds \cite{Fernandez}. 
\begin{table}[]
    \centering
    \begin{tabular}{|c|c|c|c|}
 \hline
     & Algebra & $\left[ X_i,X_j\right]$ & Completely solvable  \\
    \hline
    a) & $4g_1$ &  & abelian \\
       \hline
    b) & $g_{3,1}\oplus g_1$ & $\left[X_2,X_3\right]=X_1 $ & nilpotent\\
       \hline
    d) & $g_{4,1}$ & $\left[X_2,X_4\right]=X_1 $  , $\left[X_3,X_4\right]=-X_2 $ & nilpotent \\
    \hline
    \end{tabular}
    \caption{Nonisomorphic nilpotent algebras in four dimensions from \cite{Bock}}
    \label{tab5ta}
\end{table}
These nilmanifolds are all symplectic, and they can be understood as torus bundles over torus, or equivalently, as circle bundles over torus \cite{Bock,Fernandez,Hull11,Palais,Latorre,Belegradek}.

Let us consider a Lie group given by
\begin{eqnarray}
    G = \mathbb{R} \rtimes_\mu N
\end{eqnarray}
with a lattice $\Gamma=\mathbb{Z}\rtimes_\nu \Gamma_N$, and $N$ the nilradical. Following  \cite{Bock}, we have that $\Gamma_N$ is given by $\mathbb{Z}^3$ for $N$ abelian and by $\Gamma_{3,n}(\mathbb{Z})$ for $N$ non-abelian with $\mu:\mathbb{Z}\to\mbox{Aut}(N)$ and $\nu:\mathbb{R}\to\mbox{Aut}(\Gamma_N)$.
\begin{itemize}
    \item First case: Let us consider $\Gamma_N=\mathbb{Z}^3$, we have that $\nu:\mathbb{R}\to \mbox{SL}(3,\mathbb{R})$. Let us denote by $\lambda_1,\lambda_2,\lambda_3\in\mathbb{C}$ the roots of $\nu(1)\in \mbox{SL}(3,\mathbb{R})$.

    If $\lambda_1=\lambda_2=\lambda_3=1$ and $\nu(1)$ is diagonalisable, then the nilmanifold correspond to the abelian case, a torus $T^4$. If $\nu(1)$ is not diagonalisable, we can consider $G=\mathbb{R}\rtimes_\mu \mathbb{R}^3$ and we have two option for  $\mu(k)$. If
\begin{eqnarray}
   \mu(k)= \exp{ \begin{pmatrix}
                 0 & 0 & 0 \\
                 0 & 0 & k \\
                 0 & 0 & 0
             \end{pmatrix}} = \begin{pmatrix}
                 1 & 0 & 0 \\
                 0 & 1 & k \\
                 0 & 0 & 1
             \end{pmatrix}  
\end{eqnarray}
    then the nilmanifold corresponds to a Primary Kodaira surface with $b_1=3$ (case (b) from tables \ref{tab4ta} and \ref{tab5ta}). If
    \begin{eqnarray}
      \mu(k)=  \exp{\begin{pmatrix}
                 0 & k & -\frac{k}{2} \\
                 0 & 0 & k \\
                 0 & 0 & 0
             \end{pmatrix}}= \begin{pmatrix}
                 1 & k & \frac{1}{2}(k^2-k) \\
                 0 & 1 & k \\
                 0 & 0 & 1
             \end{pmatrix}  
    \end{eqnarray}
    then it corresponds to the nilmanifold with no complex structure and $b_1=2$ (case (d) from tables \ref{tab4ta} and \ref{tab5ta}).

\item Second case: Let us consider $\Gamma_N=\Gamma_{3,k}(\mathbb{Z})$, we have that $\nu:\mathbb{Z}\to\mbox{Aut}(\Gamma_{3,k}(\mathbb{Z}))$. This automorphism induce and automorphism $\bar\nu(1)$ which lies in $\mbox{GL}(2,\mathbb{R})$ \cite{Bock}. Denote by $\tilde{\lambda}_1,\tilde{\lambda}_2$ the roots of $\bar{\nu}(1)\in \mbox{GL}(2,\mathbb{Z})$.

If $\tilde{\lambda}_1=\tilde{\lambda}_2=1$, we can assume $\bar{\nu}(1)=\begin{pmatrix}
    1 & k' \\
    0 & 1
\end{pmatrix} \in \mbox{SL}(2,\mathbb{Z})$, with $k'\in\mathbb{N}$, and $G/\Gamma_N$ is diffeomorphic to a nilmanifold. It will correspond to a Primary Kodaira surface with $b_1=3$ (case (b) from tables \ref{tab4ta} and \ref{tab5ta}) if $k'=0$ and to a nilmanifold with no complex structure and $b_1=2$ (case (d) from tables \ref{tab4ta} and \ref{tab5ta}) if $k'\neq 0$ \cite{Bock}.
\end{itemize}

Let us emphasize that, in both cases, there is a particular Primary Kodaira surface that appears when $k=1$, and it is the Kodaira--Thurston manifold. It is the earliest known example of a symplectic manifold that is not Kahler \cite{Kodaira,Latorre,Ovando}, and the corresponding structure equations are given by
\begin{equation}
    d\eta^1=d\eta^2=d\eta^3=0,\quad d\eta^4=\eta^1\wedge\eta^2
\end{equation}.

\subsubsection{Five dimensions}

Nilmanifolds in five dimensions are in correspondence with the set of non-isomorphic nilpotent Lie algebras. There are three types of decomposable algebras and six types of indecomposable algebras, as shown in table \ref{tab6ta}.
\begin{table}[]
    \centering
    \begin{tabular}{|c|c|c|c|}
 \hline
     & Algebra & $\left[ X_i,X_j\right]$ & Compl. solv.  \\
    \hline
    a) & $5g_1$ &  & abelian \\
       \hline
    b) & $g_{3,1}\oplus 2g_1$ & $\left[X_2,X_3\right]=X_1 $ & nilpotent\\
       \hline
    d) & $g_{4,1}\oplus g_1$ & $\left[X_2,X_4\right]=X_1 $  , $\left[X_3,X_4\right]=X_2 $ & nilpotent \\
    \hline
    e) & $g_{5,1}$ & $\left[X_3,X_5\right]=X_1 $  , $\left[X_4,X_5\right]=X_2 $ & nilpotent \\
    \hline
    f) & $g_{5,2}$ & $\left[X_2,X_5\right]=X_1 $  , $\left[X_3,X_5\right]=X_2 $ , $\left[X_4,X_5\right]=X_3 $  & nilpotent\\
    \hline
    g) & $g_{5,3}$ & $\left[X_2,X_4\right]=X_3 $  , $\left[X_2,X_5\right]=X_1 $ , $\left[X_4,X_5\right]=X_2 $  & nilpotent \\
    \hline
    h) & $g_{5,4}$ & $\left[X_2,X_4\right]=X_1 $  , $\left[X_3,X_5\right]=X_1 $  & nilpotent \\
    \hline
    i) & $g_{5,5}$ & $\left[X_3,X_4\right]=X_1 $  , $\left[X_2,X_5\right]=X_1 $ , $\left[X_3,X_5\right]=X_2 $  & nilpotent \\
    \hline
    j) & $g_{5,6}$ & \makecell{$\left[X_3,X_4\right]=X_1 $ , $\left[X_2,X_5\right]=X_1 $  ,\\ $\left[X_3,X_5\right]=X_2 $ , $\left[X_4,X_5\right]=X_3 $}  & nilpotent \\
    \hline
    \end{tabular}
    \caption{Nonisomorphic nilpotent algebras in five dimensions from \cite{Bock}}
    \label{tab6ta}
\end{table}

We have already mentioned that in this work, we will be especially interested in nilmanifolds in three, four, and five dimensions. In three dimensions, there is a unique nilmanifold up to isomorphisms, but that is not the case in four and five dimensions. However, we notice that nilmanifolds associated with decomposable Lie algebras related to the Heisenberg algebra will appear in the following sections. The other nilmanifolds, as well as, more general solvmanifolds will not be considered in this work, but they will be considered elsewhere in the same context.

\section{QM2-branes and type IIB gauged supergravities}\label{Sec2}

In this section, we review some known features of QM2-branes and their relation with type IIB gauged supergravities in nine dimensions, as stated in \cite{mpgm2,mpgm7,mpgm23,mpgm24}. We will also emphasize some generalizations of \cite{mpgm23,mpgm24} that will be relevant in the following sections. These QM2-branes formulated in $M_9\times T^2$, are such that the corresponding $\mbox{SU}(N)$ regularized model has a discrete supersymmetric spectrum with finite multiplicity \cite{Boulton}. The worldvolume description is known, as well as the global description in terms of bundles, which is necessary because of the presence of topological invariants. We will review the result from \cite{mpgm24}, where we found that the transformation within equivalent bundles reproduces the symmetry group of the type IIB gauge supergravities in nine dimensions known as the triplet. In contrast, the transformation between inequivalent bundles should correspond to the U-duality transformation, a subgroup of $\mbox{SL}(2,\mathbb{Z})$. We will focus on parabolic monodromies, which are of special interest for this work.

\subsection{QM2-branes}
The action of the supermembrane (M2-brane) coupled to eleven-dimensional supergravity was originally proposed in \cite{Bergshoeff,Bergshoeff3}. The light-cone gauge (LCG) Hamiltonian was found in \cite{deWit2}. However, the $\mbox{SU}(N)$ regularized model associated with this Hamiltonian has a continuous spectrum \cite{deWit7}. Therefore, it was stated that the M2-brane is unstable. Classically, these instabilities are related to the presence of string configurations, known as spikes, with vanishing energy. It means that M2-branes with different topologies or any number of membranes can be related to the same energy \cite{Nicolai}. It was claimed that these spikes are also present on the M2-brane with winding \cite{deWit3,deWit4}. 

Nevertheless, in \cite{Restuccia}, it was found that when the M2-branes have nonvanishing irreducible wrapping on a torus, the strings configurations do not generate any classical instabilities in the theory since they carry energy and are dynamical excitations. This is because of the presence of a topological restriction on the embedding maps of the theory. The corresponding $\mbox{SU}(N)$ regularized model has a discrete supersymmetric spectrum with finite multiplicity \cite{Boulton}. Since then, different extensions with the same spectral behavior have been found; see, for example, \cite{mpgm6,mpgm25,mpgm26,mpgm27,mpgm28,mpgm29}. We refer by QM2-branes to any of these formulations with discrete spectra. In this work, we will follow the approach of \cite{mpgm6}.

\subsubsection{Worldvolume description}

The light cone gauge (LCG) bosonic Hamiltonian for an M2-brane in the presence of a non-vanishing three-form background was given in \cite{deWit}, where the authors have considered a general spacetime with metric $G_{\mu\nu}$ written in a convenient form using the gauge $$G_{--}=G_{a-}=0.$$ Its supersymmetric extension on a Minkowski spacetime, $\mathbb{R}^{1,10}$, was obtained in \cite{mpgm6} (where it was also shown its consistency with $C_{\mu\nu\rho}$ constant, not necessarily zero), and it is given by 
\begin{eqnarray}\label{HCM2}
\mathcal{H}&=&\frac{1}{2(\widehat{P}_--TC_-)}\left((\widehat{P}_a-TC_a)^2+\frac{T^2}{2}(\epsilon^{uv}\partial_u X^a \partial_v X^b)^2\right) \nonumber \\
&-& T\left(\bar{\theta}\Gamma^-\Gamma_a \left\lbrace X^a,\theta \right\rbrace - C_{+-}- C_+\right), 
\end{eqnarray}
subject to the first and second-class constraints
\begin{eqnarray}
\widehat{P}_a\partial_u X^a + \widehat{P}_- \partial_u X^- + \bar{S}\partial_u \theta &\approx& 0, \\
S - (\widehat{P}_--TC_-)\Gamma^- \theta &\approx& 0 ,
\end{eqnarray}
with $T$ being the M2-brane tension and the unique free parameter of the theory, $\widehat{P}_a$ the canonical conjugate to $X^a$ and $S,\bar{S}$ are the conjugate momenta to $\bar{\theta},\theta$ (Majorana spinors in 11D), respectively. 

The embedding used in this paper is the same one used in seminal papers (see, for example, \cite{deWit2}). We are considering the light cone gauge, and hence the space-time indices $\mu,\nu,\rho=0,\dots,10$ are split according to $\mu=(+,-, a)$, where $a=1,\dots,9$ are the transverse indices to the null light coordinates (see, for example, \cite{deWit2}). The worldvolume indices are $i=0,1,2$, with $u,v=1,2$ labeling the spatial directions of the worldvolume. We are considering an embedding of the M2-brane on the complete 11D space-time. That is, $X^a(\sigma^1,\sigma^2,\tau)$ are maps from $\Sigma$, a Riemann surface of genus 1, to the target space $X^a: \Sigma \rightarrow M_{11}$.

The LCG three-form components are written according to \cite{deWit} as
\begin{equation}\label{CaLCG}
\begin{aligned}
& C_a  =  -\epsilon^{uv}\partial_uX^- \partial_vX^b C_{-ab} +\frac{1}{2}\epsilon^{uv}\partial_uX^b \partial_vX^c C_{abc} \, , \\
& C_{\pm}  =  \frac{1}{2}\epsilon^{uv}\partial_uX^a \partial_vX^b C_{\pm ab} \,, \qquad C_{+-}  =  \epsilon^{uv}\partial_uX^- \partial_vX^a C_{+-a} \,,
\end{aligned}
\end{equation}
where $C_{+-a}=0$ is fixed by gauge invariance of the three-form and $C_{\pm ab}$, $C_{abc}$ are assumed, in this work, to be nontrivial constants by background fixing. Let us notice that $X^-$ appears explicitly in the Hamiltonian through $C_a$ \cite{deWit}. Nevertheless, one may perform a canonical transformation of the Hamiltonian by performing the following change of variables \cite{mpgm6}
\begin{eqnarray}
   P_a=\widehat{P}_a-TC_a, \quad P_-=\widehat{P}_--TC_-, \quad S = \widehat{S}, \quad \widehat{X}^a=X^a, \quad \widehat{X}^-=X^-, \quad \widehat{\theta}=\theta. \nonumber 
\end{eqnarray}
 We may use the residual gauge symmetry generated by the constraints to impose the gauge fixing condition $P_-=P^0_-\sqrt{w}$, with $\sqrt{w}$ a regular density on the worldvolume. We can then eliminate ($X^-,P_-^0$) as canonical variables and obtain a formulation solely in terms of ($X^a,P_a$) and ($\theta,\bar{S}$).

If we consider a compactification of the target space on a flat torus $T^2$ characterized by the Teichmuller parameter $\tau\in\mathbb{C}$ with $\mbox{Im}(\tau)> 0$ and a radius $R\in\mathbb{R}$, the embedding maps are split into the noncompact and compact sectors as follows
\begin{eqnarray*}
X^a(\sigma^1,\sigma^2,\tau)=(X^m(\sigma^1,\sigma^2,\tau),X^r(\sigma^1,\sigma^2,\tau)) , \label{Embedding}
\end{eqnarray*}
with $m=1,\dots,7$ and $r=8,9$, respectively. We may perform a Hodge decomposition on the closed, but not exact, one-forms $dX^r=dX_h^r + dA^r$, where $dX_h^r$ are the harmonic one-forms, and $dA^r$ are the exact ones. $dX_h^r$ may be written in terms of a normalized basis of harmonic one-forms $d\hat{X}^r$ as $dX_h^1+idX_h^2 = 2\pi R(l_r+m_r\tau)d\hat{X}^r$. The wrapping condition on the compact sector is given by
\begin{equation}
\label{ec2notas}
\oint_{\mathcal{C}_r} d \left(X^8 + iX^9 \right)= 2 \pi R \left(l_r + m_r \tau \right) \, \in  \mathcal{L} \,,
\end{equation}
where $\mathcal{C}_r$ denotes the homology basis on $\Sigma$, $\mathcal{L}$ is a lattice on the complex plane ($\mathbb{C}$) such that $T^2=\mathbb{C}/\mathcal{L}$ and the wrapping numbers $l_r,m_r$ defines the wrapping matrix
 \begin{eqnarray}
     \mathbb{W} = \begin{pmatrix}
        l_8 & l_9 \\
          m_8 & m_9
          \end{pmatrix}     
 \label{windingmatrix}
 \end{eqnarray}
So far, the Hamiltonian of the M2-brane on a torus can be written as
\begin{eqnarray}
H^{T^2}&=&\frac{1}{2P^0_-}\int_\Sigma d^2\sigma \sqrt{w}\left[\Big(\frac{P_m}{\sqrt{w}}\Big)^2+\Big(\frac{P_r}{\sqrt{w}}\Big)^2 + \frac{T^2}{2}\left(\left\{X^m,X^n\right\}^2 + 2\left\{X^m,X^r\right\}^2 \right. \right. \nonumber \\
&+&\left. \left. \left\{X^r,X^s\right\}^2\right)\right]- \frac{T}{2P^0_-}\int_\Sigma d^2\sigma \sqrt{w} (\bar{\theta}\Gamma_-\Gamma_r\left\{X^r,\theta\right\}-T\bar{\theta}\Gamma_-\Gamma_m\left\{X^m,\theta\right\}),\nonumber \label{HamiltonianM2}
\end{eqnarray}
where $\displaystyle \left\lbrace \bullet, \bullet \right\rbrace = \frac{\epsilon^{uv}}{\sqrt{w}}\partial_u \bullet \partial_v \bullet$.
Let us make some comments about this Hamiltonian. It can be seen that the harmonic maps on the compact sector may degenerate. Moreover, the string spikes identified of the M2-brane in $M_{11}$ and in $M_{10}\times S^1$, are also present in this torus compactification. Consequently, it has a
continuous spectrum \cite{deWit3,deWit4,deWit7}. These two problems were addressed during the last decades (see \cite{Restuccia,Restuccia3,Boulton,mpgm6}) and will be summarized in the following paragraphs by ensuring the nonvanishing feature of the determinant of the wrapping matrix.

Once the dependence on $X^-$ has been eliminated, a quantization condition on $C_{\pm}$ can be imposed. This condition corresponds to a 2-form flux condition on the target space 2-torus, whose pull-back through $X_h^r$, with $r=8,9$, generates a 2-form flux condition on the M2-brane worldvolume as follows \cite{mpgm10}
 \begin{eqnarray}\label{fluxpullback}
 \int_{T^2}C_{\pm}=\frac{1}{2} \int_{T^2}C_{\pm rs} d\widetilde{X}^r\wedge d\widetilde{X}^s = \frac{1}{2} \int_{T^2}C_{\pm rs} d\widehat{X}^r\wedge d\widehat{X}^s =c_{\pm}\int_\Sigma \widehat{F} = k_{\pm},
 \end{eqnarray}
 where $C_{\pm rs}=c_{\pm}\epsilon_{rs}$ with $c_{\pm}\in \mathbb{Z}/\{0\}$, $\widetilde{X}^r$ are local coordinates on $T^2$, $k_{\pm}=nc_{\pm}$ with $n\in \mathbb{Z}/\{0\}$ and  $\widehat{F}$ is a closed 2-form defined on $\Sigma$ such that it describes a worldvolume flux condition
 \begin{equation}\label{central charge}
   \int_{\Sigma}\widehat{F} = \frac{1}{2}\int_{\Sigma}\epsilon_{rs}d\widehat{X}^r \wedge d\widehat{X}^s =n,
  \end{equation}
  where the integer $n=\det(\mathbb{W})\ne 0$ characterizing the irreducibility of the wrapping, where $\mathbb{W}$ is the wrapping matrix. Consequently, $C_{\pm}$ is a closed two-form defined on the target space torus. Indeed, the flux condition on $T^2$ implies a flux condition on $\Sigma$ which is known as the `central charge condition'. The irreducible wrapping condition ensures that the harmonic modes are nontrivial and independent.

The Hamiltonian of the M2-brane with $C_-$ fluxes becomes
\begin{eqnarray}
H^{C_-}&=&\frac{1}{2P^0_-}\int_\Sigma d^2\sigma \sqrt{w}\left[\Big(\frac{P_m}{\sqrt{w}}\Big)^2+\Big(\frac{P_r}{\sqrt{w}}\Big)^2 + \frac{T^2}{2}\left(\left\{X^m,X^n\right\}^2 + 2(\mathcal{D}_rX^m)^2 \right. \right. \nonumber \\
&+&\left. \left. (\mathcal{F}_{rs})^2+ (\widehat{F}_{rs})^2\right)\right]- \frac{T}{2P^0_-}\int_\Sigma d^2\sigma \sqrt{w} (\bar{\theta}\Gamma_-\Gamma_r\mathcal{D}_r\theta-T\bar{\theta}\Gamma_-\Gamma_m\left\{X^m,\theta\right\}),\label{HamiltonianM2NT}
\end{eqnarray}
and 
\begin{eqnarray}\label{HamiltonianC+}
     H^{C_+} &=& H^{C_-} - 2\widehat{P}_-^0 T  \int d^2\sigma \sqrt{w}C_+,
\end{eqnarray}
only differs in a constant term \cite{mpgm6}. Interestingly,  the supermembrane on $M_9\times T^2$ with a central charge condition associated with an irreducible wrapping \cite{Restuccia}, is equivalent to the Hamiltonian of a supermembrane on $M_9^{LCG}\times T^2$ on a quantized $C_-$ background, i.e. $\mathcal{H}^{CC}=\mathcal{H}^{C_-}$. Hence, the discreteness property of the former automatically implies the discreteness of the M2-brane with $C_{\pm}$ fluxes. When $C_+\ne 0$, the spectrum is discrete and shifted by a constant value.

The degrees of freedom of the theory are $X^m,A^r,\theta$. Let us emphasize that, because of the general embedding considered, these M2-branes are not, in general, free particles in the non-compact space, but extended membranes—toroidal for the case considered here—that contain a non-lineal quartic bosonic potential and its fermionic counterpart.

On the other hand, the symplectic covariant derivative is defined as \cite{Ovalle1}
\begin{eqnarray}
   \mathcal{D}_rX^m &=&D_rX^m+\left\{ A_r,X^m\right\}, \label{symp-cov-der}
\end{eqnarray}
with $D_r$ is a covariant derivative defined as \cite{mpgm2,mpgm7} and it satisfies 
\begin{eqnarray}
    (D_8+iD_9) \, \bullet  = 2\pi R (l_r+m_r\tau)\left\lbrace \widehat{X}^r,\, \bullet \right\rbrace , \nonumber 
\end{eqnarray}
 The gauge contribution is given by 
 $\widehat{F}$ the minimal curvature related to the worldvolume flux on $\Sigma$ (\ref{fluxpullback}) and 
\begin{eqnarray}
  \mathcal{F}_{rs}&=& D_rA_s-D_sA_r+\left\{ A_r,A_s\right\}, \label{Fsymp}
\end{eqnarray}
corresponds to a symplectic curvature associated with the one-form connection $A_r dX^r$, where $A^r$ contains the dynamical degrees of freedom related to the exact sector of the map on $T^2$.

This Hamiltonian is subject to the local and global constraints associated with the area-preserving diffeomorphisms (APD), isomorphic to the symplectomorphisms on a two-dimensional manifold 
\begin{eqnarray}
\left\{ \frac{P_m}{\sqrt{w}} , X^m\right\} + \mathcal{D}_r\left( \frac{P_r}{\sqrt{w}}\right)+\left\lbrace \frac{\bar{S}}{\sqrt{w}},\theta \right\rbrace  &\approx& 0, \label{LocalAPD}\\
 \oint_{C_S}\left[\frac{P_m dX^m}{\sqrt{w}} + \frac{P_r (dX_h^r+dA^r)}{\sqrt{w}} + \frac{\bar{S} d\theta}{\sqrt{w}}\right] &\approx& 0, \label{GlobalAPD}
\end{eqnarray}
which appear as a residual symmetry on the theory after imposing the LCG in the covariant formulation. In fact, we have shown that M2-branes with $C_{\pm}$ fluxes are invariant under the full group of symplectomorphisms, which considers the sectors connected and not connected to the identity. Furthermore, symplectomorphisms on $T^2$ are in one-to-one correspondence to symplectomorphisms on $\Sigma$  \cite{mpgm10}. 
Classically, this Hamiltonian does not contain string-like spikes at zero cost energy that may produce instabilities \cite{mpgm}. At the quantum level, the $\mbox{SU}(N)$ regularized theory has a purely discrete spectrum since it satisfies the sufficiency criteria for discreteness found in \cite{Boulton}. The theory preserves $1/2$ of supersymmetry when considering a minimal configuration of KK or wrapping charges and $1/4$ when considering a more general state with both winding and KK \cite{mpgm6} on the interior of the moduli space. 

\subsubsection{Global description}

The M2-branes on $M_9\times T^2$ with $C_\pm$ fluxes can be formulated on twisted torus bundles with monodromy in $\mbox{SL}(2,\mathbb{Z})$ \cite{mpgm10}. 
\begin{equation}
\label{ec7notasseccion6.2}
\widetilde{T}^3_p \rightarrow E' \rightarrow \Sigma \, ,  
\end{equation}
where the base manifold is given by the Riemann surface $\Sigma$ associated with the spatial directions of the worldvolume, and the fiber is twisted torus $\widetilde{T}^3_p$, given by  
\begin{equation}
\label{ec7notasseccion6.22}
\mbox{U}(1) \rightarrow \widetilde{T}^3_p \rightarrow T^2 \, .
\end{equation}
This geometric description shows the compatibility of two different gauge structures. 
\begin{itemize}
    \item Symplectic gauge structure, which ensures the existence of a symplectic connection under symplectomorphisms. A symplectic torus bundle is defined by $E$ the total space, $F$ the fiber, which is the torus of the compact sector of the target-space $T^2$, and the base space $\Sigma$, which is also a torus. The structure group $G$ corresponds to the group of the symplectomorphism preserving the canonical symplectic two-form on $T^2$. On $\Sigma$, there exists an induced symplectic two-form, obtained from the pullback of the two-form on $T^2$ by the harmonic map from $\Sigma$ to the fiber $T^2$. We notice that the group of symplectomorphisms in $T^2$ or $\Sigma$ is isomorphic to the area-preserving diffeomorphisms. The symplectomorphisms in $T^2$ and in $\Sigma$ define the isotopic classes with a group structure $\Pi_0(G)$, in the case under consideration $\mbox{SL}(2,\mathbb{Z})$.

The action of $G$ on the fiber produces an action on the homology and cohomology classes of $T^2$. It reduces to an action of $\Pi_0(G)$. Besides, there is a homomorphism 
\begin{eqnarray}\label{monodromy}
\mathcal{M}_G:\Pi_1(\Sigma)\rightarrow \Pi_0(\mbox{Symp}(T^2))=\mbox{SL}(2,\mathbb{Z}).
\end{eqnarray}
The subgroup of $\mbox{SL}(2,\mathbb{Z})$ determined by the homomorphism is called the monodromy of the formulation. Each homomorphism defines a linear representation acting on the first homology group in $T^2$, $\mbox{H}_1(T^2)$. This homomorphism gives to each homology and cohomology group on the bundle the structure of $\mathbb{Z}(\left[\Pi_1(\Sigma) \right])$-module. Given a monodromy, \cite{Kahn} established the existence of a one-to-one correspondence between the equivalent classes of symplectic torus bundles, induced by the module structure $\mathbb{Z}_\rho^2$ on $\mbox{H}_1(T^2)$, and the elements of $\mbox{H}^2(\Sigma, Z_\rho^2)$, the second cohomology group of $\Sigma$ with coefficients $\mathbb{Z}_{\rho}^2$. They classify the symplectic torus bundles for a given monodromy in terms of the characteristic class. In the case of a symplectic torus bundle with base a torus $\Sigma$, the classification in terms of these characteristic classes is equivalent to the classification in terms of coinvariant classes of the monodromy subgroup, acting on ($p,q$) charges. We denote the coinvariant classes simply as coinvariants. Hence, the symplectic torus bundles, with a torus as the base manifold, are classified for a given monodromy according to the inequivalent coinvariants \cite{mpgm7,mpgm2}. 
\item Nontrivial $\mbox{U}(1)$ principal bundle related to the 2-form flux condition on $\Sigma$ due to the flux condition on the target space.
\end{itemize}

In \cite{mpgm10}, it was proved that the symplectic structure and the \mbox{U}(1) principal bundle are related and generate a twisted torus bundle.

The monodromy of the symplectic torus bundle is contained in $\mbox{SL}(2,\mathbb{Z})$. Consequently, it can be trivial, or it can be contained in one of the subgroups of $\mbox{SL}(2,\mathbb{Z})$.

\paragraph{Trivial monodromy}
The Hamiltonian of the QM2-branes with trivial monodromy is invariant under two inequivalent $\mbox{SL}(2,\mathbb{Z})$ symmetries \cite{mpgm3}. One is associated with the target torus and was denoted as $\mbox{SL}(2,\mathbb{Z})_{T^2}$, while the other is associated with the base manifold and denoted as $\mbox{SL}(2,\mathbb{Z})_{\Sigma}$. These symmetry groups correspond to the area-preserving diffeomorphisms, or equivalently, symplectomorphisms, on $T^2$ and $\Sigma$, respectively \cite{mpgm10}. The flux condition \eqref{fluxpullback} ensures a one-to-one correspondence of symplectomorphisms on $T^2$ and $\Sigma$, also assumed to be a 2-torus.

 The mass operator was given in \cite{mpgm19,mpgm23}. The KK and wrapping terms were first obtained in \cite{Schwarz6}, and in \cite{mpgm23}, it is shown that they are strictly related to the QM2-brane on $M_9\times T^2$ with a flux condition \eqref{fluxpullback} associated with the irreducibility of the wrapping. Double-dimensional reduction of this mass operator reproduces the full mass operator of the type IIB $\mbox{SL}(2,\mathbb{Z})$ ($p,q$)-string winding on a circle as it is shown in \cite{mpgm23}. Its low-energy limit, together with the contribution of the case without fluxes, corresponds to the unique type II maximal supergravity in nine dimensions.
 
\paragraph{Nontrivial monodromy}

Let us consider that the monodromy on the fiber is not trivial, but given by
\begin{eqnarray}\label{monodromyfiber}
    \mathcal{M}_G=\begin{pmatrix} \mathcal{M}_{11} & \mathcal{M}_{12}\\
                        \mathcal{M}_{21} & \mathcal{M}_{22}
 \end{pmatrix}^{(\alpha+\beta)} \in \mbox{SL}(2,\mathbb{Z}),
\end{eqnarray}
where ($\alpha,\beta$) are the integers characterizing the elements of $\Pi_1(\Sigma)$ and specific values of $\mathcal{M}_{ij}$, with $i,j=1,2$ will lead to parabolic, elliptic, or hyperbolic monodromies according to its trace. 

The induced transformation on $\Sigma$, also called induced monodromy on $\Sigma$, is given by
\begin{eqnarray}
   \mathcal{M}_G^*= \Omega^{-1}\mathcal{M}_G(\alpha,\beta)\Omega = \begin{pmatrix} \mathcal{M}_{11} & -\mathcal{M}_{12}\\
                        -\mathcal{M}_{21} & \mathcal{M}_{22}
 \end{pmatrix}^{(\alpha+\beta)},. \label{inducedmonodromy}
\end{eqnarray}
with \begin{eqnarray}\label{Omega}
    \Omega=\begin{pmatrix}
    -1 & 0  \\
    0 & 1
\end{pmatrix}
\end{eqnarray}

The M2-brane sectors with central charges are invariant under the $\mbox{SL}(2,\mathbb{Z})$ symmetry transformations on $\Sigma$ and $T^2$, restricted by the monodromy subgroup \cite{mpgm3,mpgm23}.

The mass operator was given in \cite{mpgm23}. The double-dimensional reduction for parabolic monodromies given by $\mathcal{M}_p=\begin{pmatrix}
    1 & k \\
    0 & 1
\end{pmatrix}$
was done in \cite{mpgm23} for $k=1$. It leads to what the authors called $q$-strings. In this work, we will consider more general parabolic monodromies with $k>1$.
%%%%%%%%%%%%%%%%%%%%%%%%%%%%%%%%

\subsubsection{On the equivalence classes.}

Inequivalent twisted torus bundles are classified according to the coinvariant classes, briefly coinvariants, for a given monodromy \cite{mpgm3,mpgm2,mpgm7,mpgm10,mpgm24}. The coinvariants related to the fiber and the base manifold are given by 
 \begin{eqnarray}
      C_F &=& \left\lbrace Q + \mathcal{M}_g\widehat{Q}-\widehat{Q}\right\rbrace, \label{C_F}\\
      C_B &=& \left\lbrace W + \mathcal{M}_g^*\widehat{W}-\widehat{W}\right\rbrace, \label{CB}
 \end{eqnarray}
respectively, where $Q = \begin{pmatrix}
             p  \\
          q 
    \end{pmatrix}\in \mbox{H}_1(T^2)$ with $p,q\in\mathbb{Z}$ are KK charges, $W = \begin{pmatrix}
         l_1  \\
          m_1 
     \end{pmatrix}\in H^1(\Sigma)$ with $l_1,m_1\in \mathbb{Z}$ are wrapping charges, $\widehat{Q}$ and $\widehat{W}$ correspond to arbitrary charges and $\mathcal{M}_g$ is the monodromy subgroup (parabolic, elliptic or hyperbolic). Given $\mathbb{W}$ as in (\ref{windingmatrix}), we will consider the class of matrices given by 
     \begin{eqnarray}
    \mathbb{W} = \begin{pmatrix}
        l_1  & l_2 \\
         m_1 & m_2
    \end{pmatrix}\begin{pmatrix}
        1  & \frac{\lambda}{m} \\
         0 & 1
    \end{pmatrix}    
     \end{eqnarray}
     with $\lambda,m \in\mathbb{Z}$ and $l_1=ml_1'$ and $m_1=mm_1'$ with $l_1',m_1'$ relatively primes. These are the most general matrices with $W$ as the first column and $\det(\mathbb{W})=n$. We use the first column of $\mathbb{W}$ in the definition of $C_B$, but we could have used the second column also. The following reasoning is also valid in both cases. 

     As formerly discussed in \cite{mpgm7}, if the monodromy is trivial, the coinvariants contain only one element, but for a nontrivial monodromy class, the coinvariants associated with the base and the fiber contain an equivalence class of KK and wrapping charges, respectively, related to the same bundle. Nevertheless, we demonstrated that the Hamiltonian is consistently defined on these coinvariants for parabolic, elliptic, or hyperbolic monodromies \cite{mpgm23,mpgm24}.

As shown in \cite{mpgm24}, the mass operator of the M2-brane with fluxes and monodromy is invariant if
\begin{eqnarray}
     Q'&=& \Lambda Q, \label{transQ1} \\
     \mathbb{W}' &=& \Lambda^*\mathbb{W}, \label{transW1}\\
       \tau' &=& \frac{a\tau +b}{c\tau + d},\label{transtau} \\
       R' &=& R\abs{c\tau+d}, \label{transR}\\
       A &=& Ae^{-\varphi}, \\
       \Gamma &=& \Gamma e^{-i\varphi}
       \label{transtau1}
\end{eqnarray}
where the $\Lambda$ matrices given by
\begin{eqnarray}  
     \Lambda &=& \begin{pmatrix}
         a & b  \\
         c & d \end{pmatrix} = \begin{pmatrix} 
        a  & \frac{1}{q}(p'-ap) \\
         \frac{1}{p'}(aq'-q) & \frac{1}{q}(q'-\frac{apq'}{p'} + \frac{qp}{p'})
\end{pmatrix},  \label{lambda_SL(2,Q)}\\
     \Lambda^* &=&\Omega^{-1}\Lambda \Omega
\end{eqnarray}
with $\Omega$ given by \eqref{Omega} and 
\begin{eqnarray}
    a = \frac{\pm Tr(\Lambda)qp'-q'p'-qp}{qp'-pq'}, \label{Aoflambda}
\end{eqnarray} 
and $p',q'$ are the elements of $Q'$. It is worth mentioning that $\Lambda$ is not an arbitrary matrix of $\mbox{SL}(2,\mathbb{R})$, in the sense that satisfies the quantization of charges (\ref{transQ1}). That is, integer KK charges $p,q$, are mapped to $p',q'$ integer charges. 

The transformation (\ref{transQ1}) maps every coinvariant for a given monodromy onto the same coinvariant. Therefore, it maps any element of a given coinvariant onto the same coinvariant
\begin{eqnarray}\label{Paraboliccoinv}
     Q&\xrightarrow[]{\Lambda}& C_F,
\end{eqnarray}
where $\abs{Tr(\Lambda)}=2$ for parabolic monodromies, $\abs{Tr(\Lambda)}<2$ for elliptic monodromies, and $\abs{Tr(\Lambda)}>2$ for hyperbolic monodromies. Consequently, the value of $a$ in (\ref{lambda_SL(2,Q)}) given by (\ref{Aoflambda}), will depend not only on the initial and final elements but also on the monodromy subgroup. These $\Lambda$ matrices are, in general, conjugated to the one-parameter subgroups of $\mbox{SL}(2,\mathbb{R})$. Indeed, $\Lambda = U_g g U_g^{-1}$ where $U_g \in \mbox{SL}(2,\mathbb{R})$ and $g\in\left\lbrace \mathbb{R}, SO(2),SO(1,1)^+\right\rbrace$ for parabolic, elliptic, or hyperbolic monodromies, respectively. In \cite{mpgm24} we interpreted this symmetry as the origin of the symmetry group of the type IIB gauge supergravity in 9D. 

Consequently, the mass operator of the M2-brane with fluxes is invariant on the equivalence classes of charges given by the coinvariants for a given parabolic, elliptic, or hyperbolic monodromy.
     
     However, some caveats are worth mentioning. First, in \cite{mpgm23,mpgm24}, we limited our discussion to the coinvariants of the fiber associated with KK charges. In the following sections, we will extend it to the coinvariants of the base associated with the windings. Second, the parabolic, elliptic, and hyperbolic monodromies considered in \cite{mpgm23,mpgm24} were not completely general. In the following sections, we will generalize the case corresponding to parabolic monodromies.

\subsection{Link with type IIB gauged supergravities in nine dimensions}

Type II supergravities with 32 supercharges in nine dimensions were classified in \cite{Bergshoeff5}. See also \cite{Hull4, Ortin}. Besides the unique maximal supergravity, there are four type IIA deformations and four type IIB deformations. Those gauge supergravities may be obtained via a generalized Scherck--Schwarz from 10 dimensions \cite{Samtleben}, or via the embedding tensor formalism in 9 dimensions.

In \cite{mpgm24,mpgm7,mpgm2} it was shown that QM2-branes with monodromy are associated with type IIB gauged supergravities in nine dimensions. In particular, with the case known as the triplet, where the symmetry groups are given by the one-parameter subgroups of $\mbox{SL}(2,\mathbb{R})$. 

QM2-branes with monodromy are classified in terms of coinvariants. The Hamiltonian can be consistently formulated in terms of these coinvariants \cite{mpgm23,mpgm24}. The transformation within equivalent classes of bundles for a given monodromy is in correspondence with the symmetry group of the type IIB gauged supergravities in nine dimensions. On the other hand, the transformation between inequivalent bundles is related to the U-duality transformation, a subgroup of $\mbox{SL}(2,\mathbb{Z})$ \cite{mpgm24}. 

The first indicator pointing at a connection between QM2-branes and type IIB gauged supergravities is given by the relation between the gauge vector of type II supergravities in nine dimensions and the symplectic connection of the QM2-brane. The twisted torus bundle description of the QM2-branes, shows the compatibility of two different gauge structures: The symplectic and the nontrivial $\mbox{U}(1)$. 
As shown in \cite{mpgm24}, the transformation of the gauge vector of type II supergravities in nine dimensions is reminiscent from the transformation of the symplectic connection of QM2-branes under symplectomorphisms (connected and not connected with the identity). Let us emphasize that, although we are restricting ourselves in this work to type IIB gauged supergravities in nine dimensions, the relation between the transformation of the gauge vector and the symplectic connection is valid for type IIA and type IIB supergravities, as was already suggested in \cite{mpgm24}.

The $\mbox{SL}(2,\mathbb{R})$ transformation \eqref{lambda_SL(2,Q)}, maps coinvariants onto coinvariants \cite{mpgm24}. Indeed, the coinvariant can be generated as an orbit of charges when specific matrices $\Lambda\in\mbox{SL}(2,\mathbb{R})$ act on them
\begin{eqnarray}
  C_F = \Lambda \mathcal{Q}
\end{eqnarray}
where $C_F= \begin{pmatrix}
     p'  \\
     q'
\end{pmatrix}$ and $\mathcal{Q}= \begin{pmatrix}
         p  \\
     q
\end{pmatrix}$.
Consequently, it preserves the quantization of charges.
However, the $\Lambda$ matrices generally do not form a group. Nevertheless, they are matrices of $\mbox{SL}(2,\mathbb{R})$, so they are conjugated to one of the one-parameter subgroups of $\mbox{SL}(2,\mathbb{R})$. As shown in \cite{mpgm24}, they are conjugated to $\mathbb{R}$, $\mbox{SO}(2)$ and $\mbox{SO}(1,1)$ for parabolic, elliptic and hyperbolic monodromies, respectively.

In the following, we will restrict ourselves to parabolic monodromies, which are the ones we are interested in this work. In this work, we will consider more general parabolic matrices than the ones considered in \cite{mpgm23,mpgm24}. The generalizations are related to the role of the torsion, as will be explained in the following sections. In order to see other monodromies, please check \cite{mpgm24}.

\paragraph{Parabolic case}
It is generated by the abelian parabolic subgroup
\begin{eqnarray}\label{parabolic_monodromy}
     \mathcal{M}_p = \begin{pmatrix}
        1 & 1 \\
         0 & 1
    \end{pmatrix}^{(\alpha+\beta)}.
\end{eqnarray}
This parabolic representation contains the infinite inequivalent conjugate classes of parabolic monodromy
\begin{eqnarray}
   \mathcal{M}_p=
   \begin{pmatrix}
        1 & k  \\
        0 & 1
    \end{pmatrix}, \label{parabolicrep}
\end{eqnarray}
with $k=(\alpha+\beta) \in\mathbb{Z}$. The coinvariants of the fiber (\ref{C_F}) are given by
\begin{eqnarray}
     C_F &=& \begin{pmatrix} 
         p +k\widehat{q}  \\
          q 
     \end{pmatrix} = \begin{pmatrix}
            \mathbb{Z}_k  \\
          q 
     \end{pmatrix}. \label{CFP}
\end{eqnarray}
Consequently, from (\ref{Aoflambda}) we get
\begin{eqnarray}
         a=  1\, , \, \mbox{ for } \, \, (\mbox{Tr}(\Lambda))=2 
\end{eqnarray}
 and from (\ref{lambda_SL(2,Q)}) the corresponding $\Lambda$ matrix that map coinvariants onto coinvariants are given by
 \begin{eqnarray}
     \Lambda_{k,q,\mathbb{Z}} &=& \begin{pmatrix}
         1 & \frac{k\mathbb{Z}}{q}  \\
         0 & 1
     \end{pmatrix} \label{LambdakqZ}
 \end{eqnarray}
It is easy to see that these matrices form a group. This is a parabolic subgroup and this is the case that was previously studied in \cite{mpgm23,mpgm24} for $k=1$. In this work, we will generalize this analysis to $k>1$ which corresponds to torsion in the homology of order greater than one.

The previous transformation maps the coinvariant onto the same coinvariant. Therefore, is a transformation within equivalent bundles, and it reproduces the symmetry group of the corresponding type IIB gauge supergravity in nine dimensions \cite{mpgm24} There is also a transformation between inequivalent bundles. This is the transformation associated with U-duality as a subgroup of $\mbox{SL}(2,\mathbb{Z})$ \cite{mpgm24}.

Let us identify the transformations that relate inequivalent classes of M2-brane twisted torus bundles with monodromy. This is equivalent to determining the transformation that relates the different coinvariant classes associated with $\mathcal{M}_g$. It turns out that this transformation is a symmetry of the formulation if we consider the corresponding transformation on the moduli and the charges (See \cite{mpgm23} for the parabolic case with $k=1$). If the monodromy is trivial, each pair of charges ($p,q$) represents a coinvariant and the symmetry of the formulation is $\mbox{SL}(2,\mathbb{Z})$ as determined by \cite{Schwarz6}. For a nontrivial monodromy, the space of ($p,q$) points is distributed in terms of disjoint coinvariants associated with $\mathcal{M}_g$, and the M2-brane is a theory on the module of $\mathcal{M}_g$-coinvariants.

As found in \cite{mpgm23} it can be seen that the transformation between coinvariants is given by
\begin{eqnarray}
    C_{F_1}\xrightarrow[]{\Lambda_{1}^{-1}} 
    \begin{pmatrix}
        1 \\ q_1
    \end{pmatrix}
 \xrightarrow[]{\mathcal{M}_\beta} \begin{pmatrix}
     1 \\ q_2
 \end{pmatrix}
  \xrightarrow[]{\Lambda_{2}} C_{F_2}, \label{Transfk=1}
\end{eqnarray}
where
\begin{eqnarray}
    \mathcal{M}_\beta= \begin{pmatrix}
        1 & 0 \\
        1 & 1
    \end{pmatrix}^\beta, \label{Mbsubgroup}
\end{eqnarray}
with the corresponding allowed values for $\beta$ in each case, will be isomorphic to a subgroup of $\mbox{SL}(2,\mathbb{Z})$.

Let us emphasize that this transformation maps integer charges into integer charges. The particular symmetry transformations allowed between different coinvariant classes for the monodromies contained on $\mbox{SL}(2,\mathbb{Z})$ were given in \cite{mpgm24}. Now we will review the case parabolic monodromies with $k=1$, in order to generalize for higher values of $k$ in the subsequent sections.

For parabolic monodromies with $k=1$, the inequivalent classes of twisted torus bundles are given by the coinvariants (\ref{CFP}). We have $\mathbb{Z}$ coinvariants classified by the value of $q$ and  $\mathbb{Z}$ elements within each coinvariant. In this case $q\in \mathbb{Z}$ and consequently $\beta\in\mathbb{Z}$. The trivial class will correspond with $q=0$. It can be seen that the transformation between inequivalent coinvariants (\ref{Mbsubgroup}) with $\beta\in\mathbb{Z}$ corresponds to the generator of the parabolic subgroup of $\mbox{SL}(2,\mathbb{Z})$
\begin{eqnarray}
    \mathcal{M}_\beta= \left\lbrace 
    \left(\begin{array}{cc}
        1 & 0  \\ 
        0 & 1
    \end{array}\right), \left(\begin{array}{cc}
        1 & 0  \\ 
        1 & 1
    \end{array}\right), \dots, \left(\begin{array}{cc}
        1 & 0  \\ 
        \mathbb{Z} & 1
    \end{array}\right)  \right\rbrace
\end{eqnarray}

The transformation between inequivalent coinvariants is realized as a discrete symmetry of the QM2-brane theory. It corresponds to the quantization of the $\mbox{SL}(2,\mathbb{R})$ to subgroups of the $\mbox{SL}(2,\mathbb{Z})$.

 Consequently, by taking into account the corresponding transformation on the moduli and the charges, we can define a transformation 
 related to $\widetilde{\Lambda}=\Lambda_{2}\mathcal{M}_\beta {\Lambda_1}^{-1}$, as described in \cite{mpgm23}, that leaves invariant the M2-brane mass operator. It can be interpreted as a duality between inequivalent classes of M2-brane twisted torus bundles for a given monodromy.
 \begin{table}
\label{table}
\begin{tabular}{|c|c|c|c|c|}
     \hline
    Monodromy & $\mathcal{M}_g\subset \mbox{SL}(2,\mathbb{Z})$ & Gauge Group  & Quantum restriction & U-duality group  \\
 \hline
    Trivial & $\mathbb{I}$ & - & - & $\mbox{SL}(2,\mathbb{Z})$ \\
    \hline
    Parabolic & $\mathcal{M}_p$ & $\mathbb{R}$ & $\mathbb{Z}$ & $\mathbb{Z}$ \\
    \hline
       Elliptic & $\mathcal{M}_{\mathbb{Z}_3}$ & $\mbox{SO}(2)$ & $\mathbb{Z}_3$ & $\mathbb{Z}_3$ \\
        \hline
      Elliptic & $\mathcal{M}_{\mathbb{Z}_4}$ &  $\mbox{SO}(2)$ & $\mathbb{Z}_4$ & $\mathbb{Z}_5$\\ 
        \hline
      Elliptic & $\mathcal{M}_{\mathbb{Z}_6}$ &  $\mbox{SO}(2)$ & $\mathbb{Z}_6$ & $\mathbb{Z}_9$\\
        \hline
      Hiperbolic & $\mathcal{M}_{n}$ &  $\mbox{SO}(1,1)$ & -  & $\mathbb{Z}_{(2n-7)}$\\ 
      \hline
\end{tabular}    
\caption{Gauge group, discrete values of the mass parameters and U-duality groups of type IIB gauged supergravities from M2-brane twisted torus bundles}
\label{tab7ma}
\end{table}

Although we have only emphasized the case related to parabolic monodromies with $k=1$, elliptic and hyperbolic monodromies were analyzed in \cite{mpgm24}. We have that there are equivalence classes of KK charges for a given nontrivial monodromy contained in $\mbox{SL}(2,\mathbb{Z})$. There are a finite number of equivalence classes for elliptic monodromies, while for parabolic and hyperbolic monodromies, there are infinite equivalence classes. The transformation within equivalent bundles reproduces the symmetry group of type IIB gauge supergravities in nine dimensions, as shown in Table \ref{tab7ma}. Moreover, the explicit identification of the elements of the coinvariants is relevant to understanding the allowed discrete values of the parameters in the low energy from the UV description \cite{mpgm24}. The latter is not only dictated because of the quantization of charges as previously conjectured in \cite{Hull4,Hull8}, but it is a consequence of the twisted torus bundle structure of the M2-brane described in terms of coinvariants.

\section{Nilmanifolds and QM2-branes}\label{Sec3}

In this section, we will identify certain nilmanifolds in three, four, and five dimensions, with torsion in the homology, that are contained in the global description of the QM2-branes. The corresponding globally well-defined basis of one-forms satisfying Maurer--Cartan equations can be written in terms of the degrees of freedom of the theory in each case. We will see that the torsion cycles are associated with the fiber of the twisted torus bundles, which is the torus of the compact sector of the target space. The existence of more general solvmanifolds in three dimensions within the M2-brane bundle description was claimed in \cite{mpgm10}. The identification of these solvmanifolds, as well as solvmanifolds in four and five dimensions within the M2-brane bundle description, will be discussed elsewhere.

  It will be useful to denote $(\sigma^1,\sigma^2)$ as coordinates on $\Sigma$, the base manifold, and $(\widetilde{X}^1,\widetilde{X}^2)$ are coordinates on $T^2$, the fiber. This notation is in correspondence with the previous sections.

\subsection{Three dimensions}

We know that nonisomorphic solvmanifolds in three dimensions are given by table \ref{tab2da}. From section \ref{Sec1}, we know that the Mostow fibration of these solvmanifolds corresponds to torus bundles over a circle with monodromy in $\mbox{SL}(2,\mathbb{Z})$, and that is a twisted 3-torus. We will be especially interested in the solvmanifolds associated with nilpotent Lie groups.

From section \ref{Sec1}, we know that three-dimensional solvmanifolds are in correspondence with the solvable Lie algebras in three dimensions from table \ref{tab2da}. Consequently, except for the abelian algebra associated with a $T^3$, the only three-dimensional nilmanifold, up to isomorphisms, is the parabolic twisted 3-torus, or equivalently, Heisenberg nilmanifold.

\paragraph{Parabolic twisted tori (Heisenberg Nilmanifold):} We know from section \ref{Sec1} that the Heisenberg nilmanifold admits geometric descriptions. The first one is a torus bundle over a circle with parabolic monodromy, which corresponds to the Mostow fibrations of the Heisenberg nilmanifold. The second one is a nontrivial circle bundle over a torus. Both descriptions are manifest within the QM2-brane global description in terms of twisted torus bundles. 

The Heisenberg nilmanifold as a nontrivial $\mbox{U}(1)$ principal bundle corresponds to the fiber of the twisted torus bundle \eqref{ec7notasseccion6.2}. Indeed, as mentioned in \cite{mpgm10}, we obtain that there can be defined a set of one-forms
\begin{eqnarray}\label{HN3}
    \eta^1 &=& d\widetilde{X}^1 \label{HN3A.1}\\
    \eta^2 &=& d\widetilde{X}^2\label{HN3A.2}\\
    \eta^3 &=& dy + k\widetilde{X}^1d\widetilde{X}^2 \label{HN3A.3}, 
\end{eqnarray}
such that the Maurer--Cartan equations are given by
$$d\eta^3 = k\eta^1\wedge \eta^2.$$ The coordinate $y\in S^1$ is associated with the $\mbox{U}(1)$ of the flux condition. Let us note that the definition of the Heisenberg nilmanifold does not depend on the value of $y$. Indeed, for $y=0$ we still have a Heisenberg nilmanifold.

In this case, the integer $k$ corresponds to the first Chern number of the bundle, which is associated with the flux condition on the torus of the target-space \eqref{fluxpullback}. Consequently, this Heisenberg nilmanifold is strictly related to the M2-branes with fluxes with a nonvanishing wrapping term on the mass operator.

Nevertheless, the M2-brane with parabolic monodromy contains another Heisenberg nilmanifold. We may consider the torus of the target space, $T^2$, which resides in the fiber of \eqref{ec7notasseccion6.2}, fibered over one of the homology cycles of the base manifold with parabolic monodromy. If that is the case, the set of globally well-defined one-forms is given by
\begin{eqnarray}
    \eta^1 &=& d\sigma^1, \label{HN3B.1} \\
    \eta^2 &=& d\widetilde{X}^2, \label{HN3B.2}\\
    \eta^3 &=& d\widetilde{X}^1 + k\sigma^1d\widetilde{X}^2, \label{HN3B.3}
\end{eqnarray}
such that Maurer-Cartan equations are satisfied
$$d\eta^3=k\eta^1\wedge \eta^2.$$

We have that, in this case, the integer $k$ corresponds to the off-diagonal element in the monodromy matrix \eqref{parabolicrep}. Consequently, this Heisenberg nilmanifold is strictly related to the M2-brane with (nontrivial) parabolic monodromy. If $k=0$, we have that the algebra is abelian and the corresponding nilmanifold is 3-torus given by a torus bundle fibered over a circle.

The first Heisenberg nilmanifold given by \eqref{HN3A.1}-\eqref{HN3A.3}, is associated with the nontrivial $\mbox{U}(1)$ principal bundle description. The second Heisenberg nilmanifold given by \eqref{HN3B.1}-\eqref{HN3B.3}, is associated with the Mostow fibration of the parabolic twisted torus. In both cases, we have that there is a torsion cycle of order $k$ related to the compactification torus. Indeed, in each case, there is a 2-chain, $C_2^{\tilde{X}^1,\tilde{X}^2}$ for the $\mbox{U}(1)$ principal bundle and $C_2^{\sigma^1,\tilde{X}^2}$ for the Mostow fibration, whose boundary is homotopy equivalent to $kS^1$.

\subsection{Four dimensions}

Nilmanifolds in four dimensions are known. They are given in table \ref{tab5ta} and their corresponding nilpotent Lie algebras are given in table \ref{tab6ta}. These are symplectic nilmanifolds, whose global descriptions are given in terms of torus bundle over torus, or equivalently, by circle bundles \cite{Bock, Fernandez}.

Let us focus on nilmanifolds with $b_1=3$, which are known as Primary Kodaira Surfaces. The corresponding Lie algebra is decomposable and can be written in terms of the Heisenberg Lie algebra. These nilmanifolds can be easily identified within the QM2-brane bundle description.

Take for example the following basis of globally-defined one-forms
\begin{eqnarray}
    \eta^1 &=& d\widetilde{X}^1, \label{HN4A.1} \\
    \eta^2 &=& d\widetilde{X}^2, \label{HN4A.2}\\
    \eta^3 &=& dy + k\widetilde{X}^1d\widetilde{X}^2 , \label{HN4A.3}\\
    \eta^4 &=& d\sigma^1, \label{HN4A.4}
\end{eqnarray}
such that the Maurer--Cartan equations are given by
$$d\eta^3 = k\eta^1\wedge \eta^2.$$
This corresponds to the Heisenberg nilmanifold of the fiber, trivally fibered over one of the homology cycles of the base manifold. That is, a trivial circle bundle as in \cite{Fernandez}. It requires $k\neq 0$ in the Heisenberg nilmanifold, and these circle bundles are classified by $\mbox{H}^2(\tilde{T}_p^3,\mathbb{Z})=\mathbb{Z}\oplus \mathbb{Z} \oplus \mathbb{Z}_k$ \cite{Fernandez}. 

Let us note that, the previous construction required nontrivial fluxes in the quantm M2-brane. Nevertheless, according to \cite{Fernandez}, this Primary Kodaira surface can also be constructed as a circle bundle for $k=0$.

As happens in three dimensions, there is another Primary Kodaira Surface that can be consistently defined in terms of the in terms of the embedding maps of the theory. Consider, for example, the following basis
    \begin{eqnarray}
            \eta^1 &=& d\sigma^1, \label{HN4B.1} \\
    \eta^2 &=& d\widetilde{X}^2, \label{HN4B.2} \\
    \eta^3 &=& d\widetilde{X}^1 + k\sigma^1d\widetilde{X}^2 , \label{HN4B.3} \\
    \eta^4 &=& d\sigma^2, \label{HN4B.4} 
    \end{eqnarray}
such that the Maurer--Cartan equations are given by $$d\eta^3 = k\eta^1\wedge \eta^2.$$

This corresponds to a torus bundle over a torus with a parabolic monodromy. Clearly, for $k=0$ we recover the $T^4$ associated with the abelian algebra.

We note that there are two different four-dimensional nilmanifold structures in the M2-brane bundle. These are in correspondence with the two different Heisenberg nilmanifolds structures previously identified in three dimensions. The first nilmanifold is given by \eqref{HN4A.1}-\eqref{HN4A.4}, is associated with the Heisenberg nilmanifold as a circle bundle. The second nilmanifold given by \eqref{HN4B.1}-\eqref{HN4B.4}, is associated with the Mostow fibration of the twisted 3-torus. In both cases, the four-dimensional nilmanifold corresponds to Primary Kodaira Surfaces. 

However, it is interesting to notice that, for $k=1$, the nilmanifold coincides with the definition of the Kodaira--Thurston manifold \cite{Kodaira,Latorre}, which is the first known example of a symplectic manifold not Kahler. 

    \subsection{Five dimensions}

Nilmanifolds in five dimensions are known. They are given in table \ref{tab6ta} and they correspond with the nine nilpotent Lie algebras in five dimensions.

The twisted torus bundle that describes QM2-branes is given by \eqref{ec7notasseccion6.2}. The fiber is given by a nontrivial $\mbox{U}(1)$ principal bundle related to the flux condition on the torus of the target space, the base manifold is given by the torus of the worldvolume and there is a monodromy which is contained in $\mbox{SL}(2,\mathbb{Z})$. Let us note that the fiber is a Heisenberg nilmanifold, globally understood as a circle bundle with $c_1=n\neq 0$. 

It can be seen that this description can be identified as a five-dimensional nilmanifold whose corresponding Lie algebra is given by $g_{3,1}\oplus g_2$. Indeed, it can be seen that we can write a well-defined basis of global one-forms in terms of the embedding maps of the theory
\begin{eqnarray}
    \eta^1 &=& d\widetilde{X}^1 \\
    \eta^2 &=& d\widetilde{X}^2\\
    \eta^3 &=& dy + k\widetilde{X}^1d\widetilde{X}^2 , \\
    \eta^4 &=& d\sigma^1, \\
    \eta^5 &=& d\sigma^2
\end{eqnarray}
such that the Maurer--Cartan equations are given by
$$d\eta^3 = k\eta^1\wedge \eta^2.$$

This corresponds to the Heisenberg nilmanifold of the fiber, fibered over the torus of the worldvolume. Note that this nilmanifold requires $k\neq 0$.

As happens in three and four dimensions, it makes sense to consider the existence of other nilmanifold structures associated with the two Heisenberg nilmanifold structures in three dimensions. In terms of the degrees of freedom of the theory is
    \begin{eqnarray}
    \eta^1 &=& d\sigma^1 \\
    \eta^2 &=& d\widetilde{X}^2\\
    \eta^3 &=& d\widetilde{X}^1 + k\sigma^1d\widetilde{X}^2 , \\
    \eta^4 &=& d\sigma^2, \\
    \eta^5 &=& dy \label{y5}
    \end{eqnarray}
such that the Maurer--Cartan equations are given by
$$d\eta^3 = k\eta^1\wedge \eta^2.$$ 

However, in this case, as we have that $y\in S^1$ is not a degree of freedom in the M2-brane formulation, we can not identify the global one-form \eqref{y5} in the M2-brane description. Consequently, the five-dimensional nilmanifold associated with the Heisenberg nilmanifold as a nontrivial circle bundle is one that we can easily identify within the global description of the QM2-brane with parabolic monodromy. 

Let us emphasize that, except for the three-dimensional case, we are not fully characterizing all the nilmanifolds contained in the QM2-brane bundle description. We are identifying some nilmanifold structures that can be written in terms of the degrees of freedom of the theory. Consequently, it has to be explored if the other known nilmanifolds in four and five dimensions are or are not contained in the M2-brane twisted torus bundles with parabolic monodromies. The same occurs with more general solvmanifolds, in three, four, and five dimensions that may be associated with QM2-branes with elliptic, or hyperbolic mondromies.

\section{Torsion and QM2-branes with parabolic monodromies}\label{Sec4}

In the previous section, we have identified several nilmanifolds in low dimensions contained in the global description of the M2-brane with fluxes and parabolic monodromy. There are, at least, two different Heisenberg nilmanifolds in three dimensions, two different Primary Kodaira surfaces, and one five-dimensional nilmanifold, associated with the nilpotent Lie algebra $g_{3,1}\oplus 2g_1$. All these nilmanifolds contain torsion cycles in the homology of the compactification torus. From the point of view of the QM2-brane, the torsion cycles are due to the flux condition or to the parabolic monodromy of the bundle.

In this section, we will show explicitly the role that torsion plays in the local and global description of the theory. In particular, we will explicitly mention the contribution of a torsion of order greater than one in the equivalence classes of bundles and in the mass operator of the theory. Moreover, we will take into account the coinvariants of the base manifold. Considering these two issues is a generalization of the analysis done in \cite{mpgm23,mpgm24}.

\subsection{On the mass operator}
In \cite{mpgm23}, the mass operator of the M2-branes with fluxes and monodromy was computed. The invariance of the mass operator on the equivalence classes, given by the coinvariants, was shown in \cite{mpgm23} for parabolic monodromies and in \cite{mpgm24} for elliptic and hyperbolic monodromies. However, in those works, the computations only considered parabolic monodromies \eqref{parabolicrep} with $k=1$. Here, we will extend it to monodromies with $k>1$, associated with torsion of order greater than 1. To achieve this, we have first to generalize the definition of the map from circles onto circles from \cite{mpgm19,mpgm23}. 

The QM2-branes on a torus are invariant under two different $\mbox{SL}(2,\mathbb{Z})$ symmetries. These are related to the full group of symplectomorphisms on the spatial worldvolume $\mbox{SL}(2,\mathbb{Z})_\Sigma$; and on the target-space torus, $\mbox{SL}(2,\mathbb{Z})_{T^2}$. By a full group of symplectomorphisms, we mean connected and not connected with the identity. For nontrivial monodromies, it can be seen that these symmetries are restricted to the monodromy subgroup \cite{mpgm23,mpgm24}. 

Let us consider a QM2-brane on $M_9\times T^2$ with a parabolic monodromy given by \eqref{parabolicrep}. Let us generalize the harmonic maps from \cite{mpgm19,mpgm23} by considering a wrapping matrix given by
\begin{eqnarray}
    \widetilde{\mathbb{W}} = S \begin{pmatrix}
        \lambda_1 & 0 \\
        0 & \lambda_2
    \end{pmatrix}\label{SW}
\end{eqnarray}
such that $\displaystyle S = \begin{pmatrix}
    a & b \\
    c & d
\end{pmatrix}\in \mbox{SL}(2,\mathbb{Z})$ and $\det \widetilde{\mathbb{W}}=\lambda_1\lambda_2 = n$. As we will see, for the wrapping matrix to have integer entries, it may be enough to consider certain matrices of $S\in \mbox{SL}(2,\mathbb{R})$. 

Let us also consider, the following set of one-forms $$d\mathbb{X}^r=\mathbb{N}^r_sdX^s,$$ with $$\mathbb{N} = \mathbb{M}S,$$ where $r,s=1,2$, and $\displaystyle \mathbb{M}= \begin{pmatrix}
    1 & \mbox{Re}(\tau) \\
    0 & \mbox{Im}(\tau)
\end{pmatrix}$. Then, it can be seen that
\begin{eqnarray}
    \frac{1}{2\pi R}\oint_{C_S}  \mathbb{N}^{-1} \begin{pmatrix}
        dX^1 \\
        dX^2
    \end{pmatrix} = \oint_{C_S} \begin{pmatrix}
        \lambda_1 d\widetilde{X}^1 \\
        \lambda_2 d\widetilde{X}^2
    \end{pmatrix}
\end{eqnarray}
defines a map from circle onto circle, where $C_S$ the homology basis of $\Sigma$ and $d\widetilde{X}^r$ the $\mbox{SL}(2,\mathbb{Z})_\Sigma$ transformation, restricted to the monodromy subgroup.

It is easy to verify that the definitions considered in \cite{mpgm23,Restuccia5,mpgm19} are contained as particular cases of this more general definition of the map from circle onto circle. As we will see, by considering this generalization we can extend the formulation of previous works to parabolic monodromies with $k>1$. Consequently, it will allow us to track the contribution of torsion of orders greater than one in the local and global formulation. 

If we compute the KK term on the mass operator following \cite{mpgm19,mpgm23}, we have to consider $\mathbb{P}$, the conjugate momenta to $\mathbb{X}$. Then we obtain
\begin{eqnarray}
    M_{KK}^2=\frac{\vert t_1\tau-t_2\vert^2}{(R\mbox{Im}(\tau))^2}, \label{KKterm}
\end{eqnarray}
with 
\begin{eqnarray}
    t_1 &=& d\widehat{m}_8 - c\widehat{m}_9, \label{t1}\\
    t_2 &=& a\widehat{m}_9-b\widehat{m}_8, \label{t2}
\end{eqnarray}
where $a,b,c$, and $d$ correspond to the entries of the matrix $S$ and $\widehat{m}_8,\widehat{m}_9$ are the integers associated with the KK quantization condition as in \cite{mpgm19,mpgm23}. It can be seen that the KK term is invariant under the $\mbox{SL}(2,\mathbb{Z})_{T^2}$ symmetry restricted to the monodromy subgroup.

The KK term \eqref{KKterm} has the same expression as the cases with monodromies associated with $k=0$ (trivial monodromy) or $k=1$ from \cite{mpgm19,mpgm23}. However, the expressions for $t_1,t_2$ are more general. Indeed, when $\lambda_1=n$ and $\lambda_2=1$ we reproduce the results from \cite{mpgm23}, and if additionally $S=\mathbb{I}$, then we obtain the results from \cite{mpgm19}. 

The full mass operator also contains a wrapping term and the membrane excitations. Those were computed in \cite{mpgm23} for arbitrary monodromies. Consequently, the mass operator of the QM2-brane with parabolic monodromy is given by
\begin{eqnarray}
    M^2_{C_-}=\left( TnA_{T^2}\right)^2 + b^2\vert \tau_R^TQ\vert^2 + 2\widehat{P}_-^0H'^{C_-}, \label{massop}
\end{eqnarray}
where
\begin{eqnarray}
    \tau_R^T=\frac{1}{R\mbox{Im}(\tau)}\begin{pmatrix}
        -1 & \tau
    \end{pmatrix}\, ,\quad Q=\begin{pmatrix}
        t_1 \\ t_2
    \end{pmatrix}
\end{eqnarray}
and $T$ is the tension of the M2-brane, $A_{T}^2=\left(2\pi R\right)^2\mbox{Im}(\tau)$ is the area, $b$ is a constant with the proper dimensions, and $H'_{C_-}$ are the nonzero modes of the Hamiltonian \eqref{HamiltonianM2NT}. 

\subsection{On the inequivalent classes of bundles}
We know that the global description of QM2-branes is classified in terms of the coinvariants. In \cite{mpgm24}, we were able to extend the Hamiltonian formulation to the full coinvariants for parabolic, elliptic, and hyperbolic monodromies. Consequently, the full theory is consistently formulated in the module of $\mathcal{M}_g$-coinvariants. However, for parabolic monodromies, which is the case we are especially interested in this work, we only analyzed the case corresponding to $k=1$ in \eqref{parabolicrep}. That is a torsion of order 1. In the following, we will consider torsion of order $k>1$ such that it is easier to identify the role of torsion.

We can write the monodromy of the twisted torus bundle as in \cite{Kahn}
$$ \mathcal{M} = \begin{pmatrix}
        1-2k'k & k + 2k'k^2 \\
        -k' & 1+k'k
    \end{pmatrix}^{a+b}$$
    where $k,k'\in \mathbb{Z}\geq 0$ and $(a,b)\in\mathbb{Z}^2$ are integers related to the homology basis of $\Sigma$. Inequivalent bundles are classified by $\mbox{H}^2(\Sigma, \mathbb{Z}^2_\mathcal{M})=\mathbb{Z}_k'\oplus\mathbb{Z}_k$. There are $k'k$ bundles for $k',k\neq 0$.
    
    Notice that when $k'=0$ 
    we have that
$$ \mathcal{M} = \begin{pmatrix}
        1 & k \\
        0 & 1
    \end{pmatrix}^{a+b}$$
    and the bundles are classified by $\mbox{H}^2(\Sigma, \mathbb{Z}^2_\mathcal{M})=\mathbb{Z}\oplus\mathbb{Z}_k$. These are parabolic monodromies. It is evident that $k=0$ corresponds to trivial monodromy. Regarding nontrivial monodromies, it will be useful to consider two possibilities: $k=1$ and $k\neq 1$.
    \begin{itemize}
        \item $k=0$ \\
        This case corresponds to trivial monodromy. The coinvariants
        \begin{eqnarray}
            C_F = \begin{pmatrix}
                p \\
                q
            \end{pmatrix},
        \end{eqnarray} 
        contains only one element. Inequivalent classes of coinvariants are given by $\mbox{H}^2(\Sigma, \mathbb{Z}^2_\mathcal{M})=\mbox{H}^2(\Sigma, \mathbb{Z}^2)=\mathbb{Z}\oplus\mathbb{Z}$. This case was considered in \cite{mpgm23}. The double-dimensional reduction corresponds to type IIB $\mbox{SL}(2,\mathbb{Z})$ ($p,q$)-strings on a circle, and the low-energy limit is given by type II maximal supergravity in nine dimensions. In this case, the only nilmanifolds in three, four, and five dimensions that can be identified within the global description, are those related to the Heisenberg nilmanifold as a nontrivial circle bundle. We are excluding here the Abelian case, which is evident and will be also present.
                \item $k=1$ \\
        The coinvariants for parabolic monodromy with $k=1$ are given by 
        \begin{eqnarray}
            C_F = \begin{pmatrix}
                \mathbb{Z} \\
                q
            \end{pmatrix}.
        \end{eqnarray}
        Each equivalence class contains more than one element. Inequivalent classes of coinvariants are given by $\mbox{H}^2(\Sigma, \mathbb{Z}^2_\mathcal{M})=\mathbb{Z}$. Indeed, different values of $q$ identify inequivalent bundles, while different values of $\mathbb{Z}$ identify the elements within the same equivalence class. This case was considered in \cite{mpgm23}. The double-dimensional reduction leads to what was called $q$-strings and the low-energy limit is given by parabolic type IIB gauged supergravity in nine dimensions. As there is a nontrivial monodromy in this case, we can identify all the nilmanifolds in three, four, and five dimensions from the previous section. However, the Primary Kodaira surfaces associated with the Mostow fibrations of the Heisenberg nilmanifold will correspond to a Kodaira--Thurston manifold in this case.
        \item $k>1$\\
        The coinvariants for parabolic monodromy with $k>1$ are given by 
        \begin{eqnarray}
            C_F = \begin{pmatrix}
                \mathbb{Z}_k \\
                q
            \end{pmatrix}.
        \end{eqnarray}
        Each one of the equivalence classes for $k=1$ is split into $k$ equivalence classes. Inequivalent classes of coinvariants are given by $\mbox{H}^2(\Sigma, \mathbb{Z}^2_\mathcal{M})=\mathbb{Z}\oplus \mathbb{Z}_k$. Indeed, is not enough to give the value of $q$ as in \cite{mpgm23}, but we need to give the integer $p\in\mathbb{Z}\mod k=\mathbb{Z}_k$ to completely determine the equivalence class. Therefore, this is more general than the case considered in \cite{mpgm23}. The QM2-brane mass operator, as well as the double-dimensional reduction and the low-energy limit, will be described in the following sections. It can be seen that if we consider $k=1$ or $k=0$ we recover the corresponding coinvariants previously mentioned. All the nilmanifolds in three, four, and five dimensions that were identified in the previous section are associated with this case. In particular, if $k>1$, Primary Kodaira surfaces will not be Kodaira--Thurston manifolds in any of the cases.
    \end{itemize}
    
    We can consider the more general coinvariants with $k\in \mathbb{Z}$, such that the integer $k$ could be $0$, $1$, or $>1$. It can be seen that the transformation that maps coinvariants onto the same coinvariants is given by \eqref{LambdakqZ} such that 
    \begin{eqnarray}
    C_F=\Lambda_{k,q,\mathbb{Z}} Q.    
    \end{eqnarray}
    We note that, as in this case we have $\vert \mbox{Tr}(\Lambda) \vert = 2$, consequently, from \eqref{Aoflambda} we have that $a=1$ in \eqref{lambda_SL(2,Q)}. This is a symmetry of the mass operator of the QM2-brane with parabolic monodromy, as can be seen from \eqref{transQ1}-\eqref{transtau1}. 

Besides considering the case with $k=1$, it was mentioned in \cite{mpgm23} that only the coinvariant of the fiber was considered. Let us try to complement this analysis with the coinvariant of the base manifold \eqref{CB}. These coinvariants correspond to equivalence classes of elements of $H^1(\Sigma)$ and they are related to the wrapping matrix \cite{mpgm7}.  Moreover, in our particular case, the wrapping is restricted by the flux condition that ensures the nonvanishing feature of $n=\det\mathbb{W}$.

Let us note that the $\mbox{SL}(2,\mathbb{R})$ transformation \eqref{LambdakqZ} maps wrapping charges on the coinvariant of the base \eqref{CB} such that
    \begin{eqnarray}
        C_B = \begin{pmatrix}
            l_1-k\mathbb{Z} \\
            m_1
        \end{pmatrix} \label{CoinvBase}
    \end{eqnarray}   
if $m_1=q$. The corresponding action on the wrapping matrix leads to
\begin{eqnarray}
    \mathbb{W} ' =\Lambda^*\mathbb{W} = \begin{pmatrix}
        l_1 - k\mathbb{Z} & l_2-c_2k\mathbb{Z} \\
        m_1 & m_2
    \end{pmatrix}, \label{windinglambdastar}
\end{eqnarray}
 where $m_2=c_2 q$ and $c_2\in\mathbb{Z}$. Consequently, $\det \mathbb{W} = n \propto q$. Let us note that, if $m_1,m_2$ are not proportional to $q$, the entries of the wrapping matrix are not integers, but the determinant, which is the one that appears in the wrapping term of the mass operator, is still given by $n$. 
 %This case with non-integer entries, may have a relation, or an interpretation similar to what occurs in open string theory \cite{} or compactification on freely acting orbifolds \cite{}. 
 In this work, we will not deal with such issues because we are considering $m_1=q$ and $m_2=c_2q$. 

Let us consider a QM2-brane with a given flux \eqref{fluxpullback}  and a given parabolic monodromy \eqref{parabolicrep}. The former is characterized by $n\in\mathbb{Z}-\left\lbrace 0 \right\rbrace$, and the latter by $k\in\mathbb{Z}-\left\lbrace 0 \right\rbrace$. If we denote by $l_1',l_2',m_1',m_2'$ the entries of $\mathbb{W}'$, we have that $l_1,l_1'\in\mathbb{Z}\mod k$ and $l_2,l_2'\in\mathbb{Z}\mod c_2k$. Let us also consider that the wrapping matrix can be written as
\begin{eqnarray}
    \widetilde{\mathbb{W}}=S\begin{pmatrix}
        \lambda_1 & 0\\
        0 & \lambda_2
    \end{pmatrix}
\end{eqnarray}
such that $S\in \mbox{SL}(2,\mathbb{Z})$ and $\lambda_1\lambda_2=n$. This choice may seem arbitrary at this stage, but in the following section, we will argue that it allows us to generalize the maps from circles onto circles from \cite{mpgm23}. Let us note that if $S\in \mbox{SL}(2,\mathbb{R})$ it could also work for our purposes if the entries of the wrapping matrix are still integers. We will come back to this issue later on. 

We do not have a reason to fix the value of $c_2$, so let us consider three different possibilities:
\begin{itemize}
    \item 1st: $c_2=0$. \\
    In this case, we have that the wrapping matrix \eqref{windinglambdastar} can be written as
    \begin{eqnarray}
    \mathbb{W} ' = \begin{pmatrix}
        l_1 - k\mathbb{Z} & l_2 \\
        q & 0
    \end{pmatrix}
\end{eqnarray}
where
    $l_1\in\mathbb{Z}\mod k$, $l_2\in\mathbb{Z}$, and $\det\mathbb{W}=-l_2q$. The possible values of $q$ are given by the divisors of $n$
    \begin{eqnarray}
        q\in\mbox{Div}(n), \label{qc2cero}
    \end{eqnarray}
 for all values of $k$, as shown in Table \ref{c2cero}.
    \begin{table}[]
    \centering
    \resizebox{\textwidth}{!}{\begin{tabular}{|c||c|c|c|c|c|c|c|c|}
 \hline
   \diagbox[width=3em]{k}{n}  & 1 & 2 & 3 & 4 & 5 & 6 & 7 & $\cdots$  \\
    \hline \hline
    1 & $q=1$ & $q=1,2$ & $q=1,3$ & $q=1,2,4$ & $q=1,5$ & $q=1,2,3,6$ & $q=1,7$ & $\cdots$  \\
    \hline
        2 & $q=1$ & $q=1,2$ & $q=1,3$ & $q=1,2,4$ & $q=1,5$ & $q=1,2,3,6$ & $q=1,7$ & $\cdots$ \\
    \hline
        3 & $q=1$ & $q=1,2$ & $q=1,3$ & $q=1,2,4$ & $q=1,5$ & $q=1,2,3,6$ & $q=1,7$ & $\cdots$  \\
    \hline
    \vdots & \vdots & \vdots & \vdots & \vdots & \vdots & \vdots & \vdots & $\ddots$ \\
    \hline
    \end{tabular}}
    \caption{Possible values of $q$ given $n$ and $k$ for $c_2=0$.}
    \label{c2cero}
\end{table}
It can be seen that the coinvariants \eqref{CoinvBase} are given by
\begin{eqnarray}
    C_B = \begin{pmatrix}
        q\mathbb{Z} \\
        q
    \end{pmatrix}
\end{eqnarray}
with $q\mathbb{Z}\in\mathbb{Z}\mod k$.

If we consider the transformation between inequivalent coinvariants from \eqref{Transfbetween}, we have that the restricted values for $\beta$ are given by table \ref{c2cerobeta}.
    \begin{table}[]
    \centering
    \begin{tabular}{|c||c|c|c|c|c|c|c|c|}
 \hline
   \diagbox[width=3em]{k}{n}  & 1 & 2 & 3 & 4 & 5 & 6 & 7 & $\cdots$  \\
    \hline \hline
    1 & - & $\beta=1$ & $\beta=2$ & $\beta=1,2,3$ & $\beta=4$ & $\beta=1,2,3,4,5$ & $\beta=6$ & $\cdots$  \\
    \hline
 \vdots & \vdots & \vdots & \vdots & \vdots & \vdots & \vdots & \vdots & $\ddots$ \\
    \hline
    \end{tabular}
    \caption{Possible values of $\beta$ given $n$ and $k$ for $c_2=0$.}
    \label{c2cerobeta}
\end{table}

    \item 2nd $c_2=1$ \\
In this case, we have that the wrapping matrix \eqref{windinglambdastar} can be written as
\begin{eqnarray}
    \mathbb{W} ' = \begin{pmatrix}
        l_1 - k\mathbb{Z} & l_2-k\mathbb{Z} \\
        q & q
    \end{pmatrix}
\end{eqnarray}
where $l_1,l_2\in\mathbb{Z}\mod k$, and $\det \mathbb{W} = n = (l_1-l_2)q$. The possible values of $q$ are given by divisors of $n$ such that 
\begin{eqnarray}
q\in \mbox{lcm}(\lambda_1,  \lambda_2)\mathbb{Z}\leq n, \label{qc2uno}    
\end{eqnarray} 
and it value depend on $k$, as shown in Table \ref{c2uno}.
    \begin{table}[]
    \centering
    \begin{tabular}{|c||c|c|c|c|c|c|c|c|}
 \hline
   \diagbox[width=3em]{k}{n}  & 1 & 2 & 3 & 4 & 5 & 6 & 7 & $\cdots$  \\
    \hline\hline
    1 &  &  &  & & &  &  & $\cdots$  \\
    \hline
        2 & $q=1$ & $q=2$ & $q=3$ & $q=4$ & $q=5$ & $q=6$ & $q=7$ & $\cdots$ \\
    \hline
        3 & \vdots & $q=1,2$ & $q=3$ & $q=2,4$ & $q=5$ & $q=3,6$ & $q=7$ & $\cdots$  \\
    \hline 
    4 & \vdots & \vdots & $q=1,3$ & $q=2,4$ & $q=5$ & $q=2,3,6$ & $q=7$ & $\cdots$ \\
    \hline 
    5 & \vdots & \vdots & \vdots & $q=1,2,4$ & $q=5$ & $q=2,3,6$ & $q=7$ & $\cdots$ \\
     \hline 
    6 & \vdots & \vdots & \vdots & \vdots & $q=1,5$ & $q=2,3,6$ & $q=7$ & $\cdots$ \\
    \hline 
    7 & \vdots & \vdots & \vdots & \vdots & \vdots & $q=1,2,3,6$ & $q=7$ & $\cdots$ \\
    \hline
    8 & \vdots & \vdots & \vdots & \vdots & \vdots & \vdots & $q=1,7$ & $\cdots$ \\
    \hline
    \vdots & \vdots & \vdots & \vdots & \vdots & \vdots & \vdots & \vdots & $\ddots$ \\
    \hline
    \end{tabular}
    \caption{Possible values of $q$ given $n$ and $k$ for $c_2=1$.}
    \label{c2uno}
\end{table}
Let us note that, the case with $k=1$ is not consistent because it implies $n=0$. 

In this case, the coinvariants are given by
\begin{eqnarray}
    C_B =  \begin{pmatrix}
        a\lambda_1 \\
        q
    \end{pmatrix}
\end{eqnarray}
with $a\lambda_1\in\mathbb{Z}\mod k$.

If we consider the transformation between inequivalent coinvariants from \eqref{Transfbetween}, we have that the restricted values for $\beta$ are given by table \ref{c2unobeta}
    \begin{table}[]
    \centering
    \begin{tabular}{|c||c|c|c|c|c|c|c|c|}
 \hline
   \diagbox[width=3em]{k}{n}  & 1 & 2 & 3 & 4 & 5 & 6 & 7 & $\cdots$  \\
    \hline \hline
    1 &  &  &  & & &  &  & $\cdots$  \\
    \hline
        2 & - & - & - & - & - & - & - & $\cdots$ \\
    \hline
        3 & \vdots & $\beta=1$ & - & $\beta=2$ & - & $\beta=3$ & - & $\cdots$  \\
    \hline 
    4 & \vdots & \vdots & $\beta=2$ & $\beta=2$ & - & $\beta=1,3,4$ & - & $\cdots$ \\
    \hline 
    5 & \vdots & \vdots & \vdots & $\beta=1,2,3$ & - & $\beta=1,3,4$ & - & $\cdots$ \\
     \hline 
    6 & \vdots & \vdots & \vdots & \vdots & $\beta=4$ & $\beta=1,3,4$ & - & $\cdots$ \\
    \hline 
    7 & \vdots & \vdots & \vdots & \vdots & \vdots & $\beta=1,2,3,4,5$ & - & $\cdots$ \\
    \hline
    8 & \vdots & \vdots & \vdots & \vdots & \vdots & \vdots & $\beta=6$ & $\cdots$ \\
    \hline
    \vdots & \vdots & \vdots & \vdots & \vdots & \vdots & \vdots & \vdots & $\ddots$ \\
    \hline
    \end{tabular}
    \caption{Possible values of $\beta$ given $n$ and $k$ for $c_2=1$.}
    \label{c2unobeta}
\end{table}

    \item 3rd $c_2 > 1$ \\
In this case, we have that the wrapping matrix \eqref{windinglambdastar} can be written as
\begin{eqnarray}
    \mathbb{W} ' =\Lambda^*\mathbb{W} = \begin{pmatrix}
        l_1 - k\mathbb{Z} & l_2-c_2k\mathbb{Z} \\
        q & c_2q
    \end{pmatrix}
\end{eqnarray}
where $l_1\in\mathbb{Z}\mod k$, $l_2\in\mathbb{Z}\mod (c_2k)$, and $\det \mathbb{W} = n = (l_1c_2-l_2)q$. The possible values of $q$ are given by divisors of $n$ such that 
\begin{eqnarray}
 q\in \frac{n}{\mbox{GCD}(\lambda_1c_2,  \lambda_2)}\mathbb{Z}\leq n , \label{generalq} 
\end{eqnarray}
and it value depend on $k$ and $c_2$, as shown in Table \ref{c2dos}. It is easy to see that this expression for $q$ reduces to the previous cases, \eqref{qc2cero} and \eqref{qc2uno}, for $c_2=0$ or $c_2=1$, respectively.
    \begin{table}[]
    \centering
    \begin{tabular}{|c||c|c|c|c|c|}
 \hline
   \diagbox[width=3em]{k}{n}  & 1 & 2 & 3 & 4 & $\cdots$  \\
    \hline \hline
        1 & $q=1$ , $c_2\geq 2$ & \makecell{$q=1$ , $c_2\geq 3$ \\ $q=2$ , $c_2\geq 2$} & \makecell{$q=1$ , $c_2\geq 4$ \\ $q=3$ , $c_2\geq 2$} & \makecell{$q=1$ , $c_2\geq 5$ \\$q=2$ , $c_2\geq 3$\\ $q=3$ , $c_2\geq 2$} & $\cdots$ \\
    \hline
        2 & $q=1$ , $2c_2\geq 2$ & \makecell{$q=1$ , $2c_2\geq 3$ \\ $q=2$ , $2c_2\geq 2$} & \makecell{$q=1$ , $2c_2\geq 4$ \\ $q=3$ , $2c_2\geq 2$} & \makecell{$q=1$ , $2c_2\geq 5$ \\$q=2$ , $2c_2\geq 3$\\ $q=3$ , $2c_2\geq 2$}  & $\cdots$  \\
    \hline 
    3 & $q=1$ , $3c_2\geq 2$ & \makecell{$q=1$ , $3c_2\geq 3$ \\ $q=2$ , $3c_2\geq 2$} & \makecell{$q=1$ , $3c_2\geq 4$ \\ $q=3$ , $3c_2\geq 2$} & \makecell{$q=1$ , $3c_2\geq 5$ \\$q=2$ , $3c_2\geq 3$\\ $q=3$ , $3c_2\geq 2$}  & $\cdots$ \\
    \hline 
    \vdots & \vdots & \vdots & \vdots & \vdots &  $\ddots$ \\
    \hline
    \end{tabular}
    \caption{Possible values of $q$ given $n$ and $k$ for $c_2>1$.}
    \label{c2dos}
\end{table}

In this case, the coinvariants \eqref{CoinvBase} are given by
\begin{eqnarray}
    C_B  =  \begin{pmatrix}
        a\lambda_1 \\
        q
    \end{pmatrix}
\end{eqnarray}
with $a\lambda_1\in\mathbb{Z}\mod k$. 

If we consider the transformation between inequivalent coinvariants from \eqref{Transfbetween}, we have that the restricted valued of $\beta$ are given in table \ref{c2dosbeta}.
 \begin{table}[]
    \centering
    \begin{tabular}{|c||c|c|c|c|c|}
 \hline
   \diagbox[width=3em]{k}{n}  & 1 & 2 & 3 & 4  & $\cdots$  \\
    \hline \hline
        1 & - , $c_2\geq 2$ & $\beta=1$ , $c_2\geq 3$ & $\beta=2$ , $c_2\geq 4$  & \makecell{$\beta=1$ , $c_2\geq 3$ \\ $\beta=2$ , $c_2\geq 5$}   & $\cdots$ \\
    \hline
        2 & \vdots & $\beta=1$ , $c_2\geq 2$ & $\beta=2$ , $c_2\geq 2$ & \makecell{$\beta=1$ , $c_2\geq 2$\\ $\beta=2$ , $c_2\geq 3$}  & $\cdots$  \\
    \hline 
    3 & \vdots & \vdots & \vdots & $\beta=1,2$ , $c_2\geq 2$  & $\cdots$ \\
    \hline 
    \vdots & \vdots & \vdots & \vdots & \vdots  & $\ddots$ \\
    \hline
    \end{tabular}
    \caption{Possible values of $\beta$ given $n$ and $k$ for $c_2>1$.}
    \label{c2dosbeta}
\end{table}
 \end{itemize}
So far we have that the possible values of $c_2$ are restricted to arbitrary integers. It can be seen that the matrix $S$ from \eqref{SW} can be written as
\begin{eqnarray}
        S= \begin{pmatrix}
             a & \pm 1 \\ \mp1 & 0
         \end{pmatrix}\, ,  \quad \mbox{ 
 for  } c_2=0\label{Sc2cero}
    \end{eqnarray} 
     and
    \begin{eqnarray}
        S= \begin{pmatrix}
             a & \frac{\lambda_1}{q\lambda_2}(aqc_2-\lambda_2) \\ \frac{q}{\lambda_1} & \frac{qc_2}{\lambda_2}
         \end{pmatrix}\, ,  \quad \mbox{ 
 for  } c_2\geq1\label{Sc2neq1}
    \end{eqnarray}
     where $q$ is given by \eqref{generalq}. It will be useful to differentiate between $c_2=1$ and $c_2\geq2$.
    
    Notice that, the only element of the matrix $S$ that can be rational and not integer is $\displaystyle \frac{\lambda_1}{q\lambda_2}(aqc_2-\lambda_2)$. However, is easy to check the wrapping matrix \eqref{SW} has integer entries in any case. Moreover, using the Bézout identity, it can be verified that there always exists integers $a$ such that $S\in \mbox{SL}(2,\mathbb{Z})$. 
    \begin{theorem}[Bézout identity]\label{Bezout}
    
    Let $d_1$ and $d_2$ be integers with $\mbox{GCD}(d_1,d_2)=C$. Then there exist $a,b\in\mathbb{Z}$ such that $$ad_1+bd_2=C.$$ Moreover, the integers of the form $ed_1+fd_2$ are exactly multiples of $C$.
    \end{theorem}
    We will emphasize it in each case in the following when we consider the different possible winding matrices associated with a given $n=\det\mathbb{W}$.

Let us notice that, once we set $\det\mathbb{W}=n$, there are different possibilities to rearrange the wrapping matrix \eqref{SW} depending on the values of $\lambda_1,\lambda_2$, and $c_2$. This will also be useful in the following section, where we will analyze the string configurations and the explicit ($p,q$)-strings associated with the QM2-branes with parabolic monodromies.

Let us first consider what happens if $n$ is a prime number, and then we consider arbitrary values of $n$.
\begin{enumerate}
    \item  If $n$ is a prime number, we have only two possibilities for the values of $\lambda_1$ and $\lambda_2$ such that $\lambda_1\lambda_2=n$.
   \begin{itemize}
       \item First case: $\lambda_1=n$, $\lambda_2=1$.
        
       Here, we have from \eqref{generalq} that $q=n$. Consequently, we can write \eqref{SW} with
       \begin{eqnarray}
           S &=& \begin{pmatrix}
              a & anc_2-1 \\
              1 & nc_2
           \end{pmatrix} \label{Snp1}\\
           \widetilde{\mathbb{W}} &=& \begin{pmatrix}
              an & anc_2-1 \\
              n & nc_2
           \end{pmatrix}\label{Wnp1}
       \end{eqnarray}
       for $c_2=0,1$ or $c_2>1$.
       
       \item Second case: $\lambda_1=1$, $\lambda_2=n$.

        In this case, we have from \eqref{generalq} that $\displaystyle q=\frac{n\mathbb{Z}}{\mbox{GCD}(c_2,n)}$. Then, we can identify different possibilities:
        \begin{itemize}
            \item If $c_2=0$, then $q=\mathbb{Z}\in \mbox{Div}(n)$, and
            \begin{eqnarray}
           S &=& \begin{pmatrix}
              a & \pm \frac{1}{q} \\
              \mp q & 0
           \end{pmatrix} \label{Snpc2cero}\\
           \widetilde{\mathbb{W}} &=& \begin{pmatrix}
              a & \pm \frac{n}{q} \\
              \mp q & 0
           \end{pmatrix}. \label{Wnpc2cero}
       \end{eqnarray}
       We notice that, as $q=\mathbb{Z}\in\mbox{Div}(n)$, the wrapping matrix has integer entries. For $q=1$, we can guarantee that the matrix $S\in \mbox{SL}(2,\mathbb{Z})$.
            \item If $c_2=1$, then $q=n$, and
            \begin{eqnarray}
           S &=& \begin{pmatrix}
              a &  \frac{a-1}{n} \\
              n & 1
           \end{pmatrix}, \label{Snpc2uno}\\
           \widetilde{\mathbb{W}} &=& \begin{pmatrix}
              a & a-1 \\
              n & n
           \end{pmatrix} \label{Wnpc2uno}
       \end{eqnarray}
       It can be seen that the existence of integers $a$ such that $S\in \mbox{SL}(2,\mathbb{Z})$ is ensured by Theorem \ref{Bezout}.
            \item If $c_2>1$ and $c_2\propto n$, we have that $\mbox{GCD}(c_2,n)=n$ and $q=1$. Then
            \begin{eqnarray}
           S &=& \begin{pmatrix}
              a &  aC-1 \\
              1 & C
           \end{pmatrix}\\
           \widetilde{\mathbb{W}} &=& \begin{pmatrix}
              a & n(aC-1) \\
              1 & Cn
           \end{pmatrix}
       \end{eqnarray}   
       where we considered $c_2=Cn$. 
       \item If $c_2>1$ and $c_2$, $n$ are co-primes, then $\mbox{GCD}(c_2,n)=1$ and $q=n$. Consequently
            \begin{eqnarray}
           S &=& \begin{pmatrix}
              a &  \frac{ac_2-1}{n} \\
              n & c_2
           \end{pmatrix}\\
           \widetilde{\mathbb{W}} &=& \begin{pmatrix}
              a & ac_2-1 \\
              n & c_2n
           \end{pmatrix}
       \end{eqnarray}
        \end{itemize}      
   \end{itemize}
   As before, using Theorem \ref{Bezout} we can ensure the existence of integers $a$ such that $S\in \mbox{SL}(2,\mathbb{Z})$.
   
    \item If $n$ is not a prime number, we have many more possibilities than before.
    \begin{itemize}
        \item First case: $\lambda_1=n$, $\lambda_2=1$.
        
      Here, we have that $q=n$ and the matrices $S$ and $\mathbb{W}$ are given by \eqref{Snp1} and \eqref{Wnp1}, respectively, for $c_2=0,1$ or $c_2>1$.
      
        \item Second case: $\lambda_1=1$, $\lambda_2=n$.
        
    Here, we have from \eqref{generalq} that $\displaystyle q=\frac{n\mathbb{Z}}{\mbox{GCD}(c_2,n)}$. Then, we can identify different cases:
        \begin{itemize}
            \item If $c_2=0$, then $q\in\mathbb{Z}\in \mbox{Div}(n)$, and matrices $S$, $\widetilde{\mathbb{W}}$ are given by \eqref{Snpc2cero} and \eqref{Wnpc2cero}, respectively.
            \item If $c_2=1$, then $q=n$, and $S$, $\widetilde{\mathbb{W}}$ are given by \eqref{Snpc2uno} and \eqref{Wnpc2uno}, respectively.
            
            \item If $c_2>1$ and $c_2,n$ are not co-primes, then $\mbox{GCD}(c_2,n)=C$ and $q=\frac{n\mathbb{Z}}{C}$ with $C\in\mbox{Div}(n)$. Consequently
              \begin{eqnarray}
           S &=& \begin{pmatrix}
              a &  \frac{ac_2-C}{n} \\
              \frac{n}{C} & c_2
           \end{pmatrix}\\
           \widetilde{\mathbb{W}} &=& \begin{pmatrix}
              a & C(ap'-1) \\
              \frac{n}{C} & np'
           \end{pmatrix}
       \end{eqnarray}      
            where we considered $c_2=Cp'$ and $n=Cq'$, with $p',q'$ co-primes. It can be seen that the existence of integers $a$ such that $S\in \mbox{SL}(2,\mathbb{Z})$ is ensured by Theorem \ref{Bezout}.
            \item If $c_2>1$ and $c_2,n$ are co-primes, then $\mbox{GCD}(c_2,n)=1$ and $q=\mathbb{Z}\in\mbox{Div}(n)$. Consequently,
            \begin{eqnarray}
           S &=& \begin{pmatrix}
              a &  \frac{ac_2-1}{n} \\
              n & c_2
           \end{pmatrix}\\
           \widetilde{\mathbb{W}} &=& \begin{pmatrix}
              a & ac_2-1 \\
              n & c_2n
           \end{pmatrix}
       \end{eqnarray} 
   \end{itemize} 
   As before, there exists $a\in\mathbb{Z}$ such that $S\in \mbox{SL}(2,\mathbb{Z})$.
        \item Third case: $\lambda_1=\frac{n}{d_1}$, $\lambda_2=\frac{n}{d_2}$, such that $d_1d_2=n$.

        Here, we can consider two different possibilities, depending if the $d_1$ and $d_2$ are equal or not. 
\begin{itemize}
    \item If $d_1=d_2=d$, then $q=d$ and $S$, $\widetilde{\mathbb{W}}$ are given by
        \begin{eqnarray}
           S &=& \begin{pmatrix}
              a &  ac_2-1 \\
              1 & c_2
           \end{pmatrix}\\
           \widetilde{\mathbb{W}} &=& \begin{pmatrix}
              \frac{an}{d} &  \frac{n(ac_2-1)}{d} \\
              \frac{n}{d} & \frac{nc_2}{d}
           \end{pmatrix}
       \end{eqnarray} 
       for $c_2=0,1$ and $c_2>1$.
    \item If $d_1\neq d_2$, then $\displaystyle q= \frac{n\mathbb{Z}}{\mbox{GCD}(d_2c_2,d_1)}$ and we can consider different cases depending on $c_2$.
\begin{itemize}
            \item When $c_2=0$, then $\mbox{GCD}(d_2c_2,d_1) = d_1$ and $\displaystyle q=\frac{n\mathbb{Z}}{d_1}=d_2\mathbb{Z}\leq n$. In this case, we have that
            \begin{eqnarray}
           S &=& \begin{pmatrix}
              a & \pm 1 \\
              \mp1 & 0
           \end{pmatrix}\\
           \widetilde{\mathbb{W}} &=& \begin{pmatrix}
               ad_2 & \pm d_1 \\
               \mp d_2 & 0 
           \end{pmatrix},
       \end{eqnarray} 
       for $\mathbb{Z}=1$.
            \item When $c_2=1$ and $d_1,d_2$ are co-primes, then $\mbox{GCD}(d_2,d_2)=1$ and $q=n$. Consequently
            \begin{eqnarray}
           S &=& \begin{pmatrix}
              a &  \frac{ad_2-1}{d_1} \\
              d_1 & d_2
           \end{pmatrix} \\
           \widetilde{\mathbb{W}} &=&  \begin{pmatrix}
               ad_2 & ad_2-1 \\
               n & n
           \end{pmatrix}
       \end{eqnarray}
       where the existence of $a$ such that $S\in \mbox{SL}(2,\mathbb{Z})$ is guaranteed by Theorem  \ref{Bezout}.
       \item When $c_2=1$ and $d_1,d_2$ are not co-primes, then $\mbox{GCD}(d_1,d_2)=C$, and $q=\frac{n\mathbb{Z}}{C}\in\mbox{Div}(n)$. Consequently
            \begin{eqnarray}
           S &=&  \begin{pmatrix}
               a &  \frac{ad_2-C}{d_1} \\
               d_1' & d_2'
           \end{pmatrix} \\
           \widetilde{\mathbb{W}} &=& \begin{pmatrix}
              ad_2 & ad_2-1 \\
              d_2d_1' & d_1d_2'
           \end{pmatrix}
       \end{eqnarray}   
       for $d_1=Cd_1'$, $d_2=Cd_2'$ and $\mathbb{Z}=1$. As before, we can ensure the existence of $a\in\mathbb{Z}$ such that $S\in \mbox{SL}(2,\mathbb{Z})$.
       
            \item When $c_2>1$, and $c_2d_2$,$d_1$ are co-primes, then $\mbox{GCD}(c_2d_2,d_1)=1$ and $q=n$. Consequently
            \begin{eqnarray}
           S &=& \begin{pmatrix}
              a &  \frac{ad_2c_2-1}{d_1} \\
              d_1 & d_2
           \end{pmatrix}\\
           \widetilde{\mathbb{W}} &=& \begin{pmatrix}
              ad_2 & ad_2c_2-1 \\
              n & n
           \end{pmatrix}
       \end{eqnarray}
       where, using Theorem \ref{Bezout}, it can be seen that there exist integers $a\in\mathbb{Z}$ such that $S\in \mbox{SL}(2,\mathbb{Z})$.
       
       \item When $c_2>1$ and $c_2d_2$, $d_1$ are not co-primes, then $\mbox{GCD}(d_1,d_2c_2)=C$, and $q=\frac{n\mathbb{Z}}{C}$. Consequently
            \begin{eqnarray}
           S &=& \begin{pmatrix}
              a &  \frac{d_2ac_2-C}{d_1} \\
              d_1' & d_2'c_2
           \end{pmatrix}\\
           \mathbb{W} &=& \begin{pmatrix}
              ad_2 & d_2ac_2-C \\
              d_1d_2' & c_2d_2d_1'
           \end{pmatrix}
       \end{eqnarray}   
    for $d_1=Cd_1'$, $d_2=Cd_2'$, and as previously, there exists $a\in\mathbb{Z}$ such that $S\in \mbox{SL}(2,\mathbb{Z})$.
   \end{itemize}
\end{itemize}
    \end{itemize}
\end{enumerate}

We have considered different possibilities for the matrices $S$ and $\mathbb{W}$ in \eqref{SW}, depending on the value of $\lambda_1,\lambda_2$ and $c_2$. Notice that, for a given $n=\det\mathbb{W}$, we have that all possible values of $\lambda_1$ and $\lambda_2$, such that $n=\lambda_1\lambda_2$ are valid. Consequently, all the coinvariants associated with $n$ are present. 

It can be seen that we could write the mass operator of the QM2-brane with parabolic monodromies in each one of the previous cases. The main difference will be on the integers $t_1$ and $t_2$, which are written in terms of the elements of the matrix $S$ as it can be seen from \eqref{t1} and \eqref{t2}.

Let us note that the generalization of the transformation within the same coinvariants from $k=1$ to $k\geq2$ is straightforward. Nevertheless, the transformation between inequivalent coinvariants does not happen in the same way. We will analyze this transformation in the following subsection.

\subsection{On the transformation between equivalence classes}

As we mentioned before, the coinvariants associated with a QM2-brane with parabolic monodromy for $k=1$ (associated with torsion of order 1) are split in $k$ coinvariants for $k\geq 2$ (associated with torsion of order greater than one). 
\begin{eqnarray}
    \left\lbrace \begin{pmatrix}
        0 \\ q
    \end{pmatrix}, \begin{pmatrix}
        1 \\ q
    \end{pmatrix}, \dots,\begin{pmatrix}
        \mathbb{Z} \\ q
    \end{pmatrix}  \right\rbrace \to \left\lbrace \begin{array}{l}
        \left\lbrace \begin{pmatrix}
        0 \\ q
    \end{pmatrix}, \begin{pmatrix}
        k \\ q
    \end{pmatrix}, \dots,\begin{pmatrix}
        k\mathbb{Z} \\ q
    \end{pmatrix}  \right\rbrace  \\
          \left\lbrace \begin{pmatrix}
        1 \\ q
    \end{pmatrix}, \begin{pmatrix}
        k+1 \\ q
    \end{pmatrix}, \dots,\begin{pmatrix}
        k\mathbb{Z} +1 \\ q
    \end{pmatrix}  \right\rbrace  \\
    \vdots \\
    \left\lbrace \begin{pmatrix}
        k-1 \\ q
    \end{pmatrix}, \begin{pmatrix}
        2k-1 \\ q
    \end{pmatrix}, \dots,\begin{pmatrix}
        k\mathbb{Z} + k-1 \\ q
    \end{pmatrix}  \right\rbrace  \\
    \end{array} \right.
\end{eqnarray}
This means that we have $k$ inequivalent coinvariants for the same value of $q$. It implies that the transformation within the same coinvariant for $k=1$ \cite{mpgm23}, is a transformation between inequivalent coinvariants with the same $q$. It also implies that the transformation between inequivalent coinvariants for $k=1$ \cite{mpgm23}, is mapped between specific coinvariants, identified by the element $\displaystyle \begin{pmatrix}
    1 \\ q
\end{pmatrix}$, for $k\geq 2$. From the previous subsection, we know that, in general, setting $n$, $k$, and $c_2$ implies that $q$ is restricted to certain divisors of $n$.

It can be seen that the general transformation between arbitrary coinvariants is given by
\begin{eqnarray}
    C_{F_1}\xrightarrow[]{\Lambda_{k,q_1,\mathbb{Z}_1}^{-1}} \begin{pmatrix}
        p_1 \\ q_1
    \end{pmatrix} \xrightarrow[]{\Lambda_{1,q_1,p_1-1}^{-1}} \begin{pmatrix}
        1 \\ q_1
    \end{pmatrix}  \xrightarrow[]{\mathcal{M}_\beta}  \left(\begin{array}{c}
     1  \\
     q_2 
\end{array}\right) \xrightarrow[]{\Lambda_{1,q_2,p_2-1}} \begin{pmatrix}
    p_2 \\ q_2
\end{pmatrix} \xrightarrow[]{\Lambda_{k,q_2,\mathbb{Z}_2}} C_{F_2}, \label{Transfbetween}
\end{eqnarray}
where $\Lambda_{k,q,\mathbb{Z}}$ and $\mathcal{M}_\beta$ are given by \eqref{LambdakqZ} and \eqref{Mbsubgroup}, respectively, and
\begin{eqnarray}
\Lambda_{1,q_1,p_1-1} &=& \begin{pmatrix}
    1 & \frac{p_1-1}{q_1} \\
    0 & 1
\end{pmatrix} , \\
    \Lambda_{1,q_2,p_2-1}  &=& \begin{pmatrix}
    1 & \frac{p_2-1}{q_2} \\
    0 & 1
\end{pmatrix} .
\end{eqnarray}
In \cite{mpgm23,mpgm24}, the coinvariants associated with parabolic monodromies for $k=1$ were characterized by $q$. There was one coinvariant for each value of $q$, and inequivalent coinvariants were isomorphic to $\mathbb{Z}$. The transformation between inequivalent classes was characterized by $\mathcal{M}_\beta$ given by expression \eqref{Mbsubgroup} as it can be seen in \eqref{Transfk=1}. There, $\beta$ was restricted to integer values. When considering torsion of order greater than one, we have that given $n$, $k$ and $c_2$, the values of $q$ are restricted to certain divisors of $n$, as shown in tables \ref{c2cero},\ref{c2uno} and \ref{c2dos}. Consequently, the values of $\beta$ are also restricted as can be seen from tables \ref{c2cerobeta}, \ref{c2unobeta} and \ref{c2dosbeta}.

Let us also note that, because of the splitting of the equivalence classes, the transformation inside the coinvariant in \cite{mpgm23,mpgm24} is a transformation between different coinvariants for $k>1$. Indeed, the $\Lambda_{1,q_1,p_1-1}$ and $\Lambda_{1,q_2,p_2-1}$ transformations in \eqref{Transfbetween} are transformations between inequivalent coinvariants. For arbitrary coinvariants, we have that the transformation is given by consecutive parabolic matrices of $\mbox{SL}(2,\mathbb{R})$ and $\mbox{SL}(2,\mathbb{Z})$.
\begin{eqnarray}
\widetilde{\Lambda}_{p_1,p_2,q_1,q_2}&=& \Lambda_{1,q_2,p_2-1}\mathcal{M}_\beta\Lambda_{1,q_1,p_1-1}^{-1}, \nonumber \\ 
&=& \begin{pmatrix}
    p_2 - \frac{q_1}{q_2}\left( p_2-1\right) & \frac{p_1\left( p_2-1\right)}{q_2} - \frac{p_2\left( p_1-1\right)}{q_1}  \\
    q_2-q_1 & p_1-\frac{q_2}{q_1}\left( p_1-1 \right)
\end{pmatrix} 
\end{eqnarray}
It can be checked that this transformation satisfies:
\begin{eqnarray}
  \widetilde{\Lambda}_{p_1,p_2,q,q} &=& \Lambda_{1,q,p_2-p_1} , \\
  \widetilde{\Lambda}_{1,1,q_1,q_2} &=& \mathcal{M}_\beta , \\
  \widetilde{\Lambda}_{p_1,p_3,q_1,q_3} &=& \widetilde{\Lambda}_{p_2,p_3,q_2,q_3}\widetilde{\Lambda}_{p_1,p_2,q_1,q_2} \\
\widetilde{\Lambda}_{p_2,p_1,q_2,q_1}\widetilde{\Lambda}_{p_1,p_2,q_1,q_2} &=& \mathbb{I}
\end{eqnarray}
The transformation between inequivalent coinvariants satisfies the same properties as the transition function on a fiber bundle. Coinvariants correspond to equivalence classes of twisted torus bundles with monodromy in $\mbox{SL}(2,\mathbb{Z})$. It would be very interesting to try to explore if this is a signal of a different geometric structure behind the global description of the QM2-brane. We leave this for subsequent works

As in \cite{mpgm23}, we may consider the full transformation, such that it is a symmetry of the mass operator. Indeed, by considering
\begin{eqnarray}
	  C_{F_{1}} &\xrightarrow[]{\widetilde{\Lambda}}&  C_{F_{2}},\quad
      \tau \rightarrow \frac{\widetilde{\Lambda}_{11}\tau + \widetilde{\Lambda}_{12}}{\widetilde{\Lambda}_{21}\tau+\widetilde{\Lambda}_{22}}  
\mathbb{W}\rightarrow \widetilde{\Lambda}^* \mathbb{W} , \nonumber \\
R&\rightarrow& R\left\vert \widetilde{\Lambda}_{21}\tau+\widetilde{\Lambda}_{22}\right\vert, \quad A\rightarrow Ae^{i\varphi_\tau},  \quad
	\Gamma \rightarrow \Gamma e^{i\varphi_\tau}, \quad \label{115}
	\end{eqnarray}
	with
    \begin{eqnarray}
	   \widetilde{\Lambda}&=& \Lambda_{k,q_2,\mathbb{Z}_2}\Lambda_{1,q_2,p_2-1}\mathcal{M}_\beta \Lambda_{1,q_1,p_1-1}^{-1}\Lambda_{k,q_1,\mathbb{Z}_1}^{-1},  \label{lambdatilde} \\
    \widetilde{\Lambda}^* &=& \Omega^{-1}\widetilde{\Lambda}\Omega , \\
    e^\varphi_\tau&=&\frac{\widetilde{\Lambda}_{21}\tau+\widetilde{\Lambda}_22}{\left\vert \widetilde{\Lambda}_{21}\tau+\widetilde{\Lambda}_22 \right\vert}
	\end{eqnarray}
and
\begin{eqnarray}
    \widetilde{\Lambda}_{11} &=& p_2+\frac{1}{q_2}\left(k\mathbb{Z}_2\beta-q_1\left( p_2-1\right)\right),\\
    \widetilde{\Lambda}_{12} &=& \frac{1}{q_1q_2}\left(-q_1(p_1+k\mathbb{Z}_1) + q_2(p_2+k\mathbb{Z}_2) -(p_1+k\mathbb{Z}_1)\beta(p_2+k\mathbb{Z}_2) \right), \\
    \widetilde{\Lambda}_{21} &=& \beta, \\
    \widetilde{\Lambda}_{22} &=& \frac{q_2-\beta(p_1+k\mathbb{Z}_1)}{q_1}\beta.
\end{eqnarray}
    It can be checked that this transformation leaves invariant the mass operator of the QM2-brane with parabolic monodromy. This can be interpreted as a duality between inequivalent classes of M2-brane twisted torus bundles with parabolic monodromies.

We have seen a generalization of the cases with $k=1$ considered in \cite{mpgm23,mpgm24}. Moreover, we have also studied the analysis of those works to the coinvariants of the base manifold. We expect that these generalizations also contribute to the string states obtained via double-dimensional reduction. In \cite{mpgm23}, we denoted $q$-strings to the parabolic ($p,q$)-strings obtained from the QM2-brane with parabolic monodromy. However, these $q$-strings correspond to monodromies with $k=1$, where there are $\mathbb{Z}$ coinvariants, $\mbox{H}^2(\Sigma,\mathbb{Z}_\rho)=\mathbb{Z}$. When $k>1$, we have that $\mbox{H}^2(\Sigma,\mathbb{Z}_\rho)=\mathbb{Z}\oplus\mathbb{Z}_k$, and we expect that each equivalence class of QM2-brane, corresponds to inequivalent parabolic ($p,q$)-strings compactified on a circle.

\section{Conclusions}\label{Sec5}

We have characterized the role that torsion plays in the formulation of QM2-branes on $M_9\times T^2$ with parabolic monodromies. This is a generalization of previous works where we considered parabolic monodromies \eqref{parabolicrep} with $k=1$. Here, we consider $k>1$ in such a way that torsion cycles of order greater than one are considered. 

The geometric description of the QM2-brane in $M_9\times T^2$, is given by twisted torus bundles. Given a symplectic torus bundle with monodromy in $\mbox{SL}(2,\mathbb{Z})$, with base manifold $\Sigma$, fiber $T^2$, and structure group $G=\mbox{Symp}(T^2)$, we may also consider the principal bundle associated with the flux condition \ref{fluxpullback}. The symplectic connection and the $\mbox{U}(1)$ connection on $\Sigma$ from \cite{mpgm10}, defines the $\mbox{U}(1)$ structure of a twisted torus. There are two ways where the torsion is manifest in the global description. The first one consists of the identification of nontrivial nilmanifolds of three, four, and five dimensions within the M2-brane bundle description. By nontrivial nilmanifolds, we are referring to those that are associated with nonabelian nilpotent Lie algebras. All the cases considered contain nonvanishing torsion in the homology. The well-defined set of one-forms can be written in terms of the degrees of freedom of the QM2-brane. The second one is characterizing the role that torsion plays in the equivalence classes of M2-brane twisted torus bundles.

In three dimensions, we identify two different Heisenberg nilmanifolds contained in the M2-brane twisted torus bundles. The first one corresponds to the fiber of the twisted torus bundle, where the Heisenberg nilmanifold is understood as a nontrivial $\mbox{U}(1)$ principal bundle with $c_1=k$. This case was discussed in \cite{mpgm10}. The second one corresponds to the Mostow fibration of the Heisenberg nilmanifold, that is, a torus bundle over a circle with parabolic monodromy characterized by the off-diagonal element, $k$. In both cases, $k\neq 0$ ensures the nonabelian feature of the Heisenberg algebra. Nevertheless, the nature of such an integer is different in each case. In the circle bundle description, $c_1=k$ is associated with the nontrivial flux condition on the compactification torus, which is responsible for the discrete supersymmetric spectrum of the theory. In the Mostow fibration, $k$ is related to the monodromy matrix, which resides in the parabolic subgroup of $\mbox{SL}(2,\mathbb{Z})$. 

In four dimensions, we identified two different structures associated with Primary Kodaira Surfaces, and related to the two Heisenberg nilmanifolds previously identified. The first one is given by a trivial circle bundle over the Heisenberg nilmanifold associated with the fiber of the twisted torus bundle. In this case, $k\neq 0$, is ensured by the flux condition on the compactification torus. The second one is given by a symplectic torus bundle over a torus with parabolic monodromy. Here, $k\neq 0$ is ensured by the nontrivial parabolic monodromy. Primary Kodaira surfaces can be consistently described as trivial circle bundles over the Heisenberg nilmanifold, or as symplectic torus bundles with parabolic monodromy. Let us emphasize that, in both cases, $k=0$ leads to a $T^4$ and $k=1$ leads to a Kodaira--Thurston manifold, which is the first known example of a symplectic manifold, not Kahler. In five dimensions, we identify the five-dimensional nilmanifold associated with the complete twisted torus bundle that globally describes the QM2-brane on $M_9\times T^2$. That is, a Heisenberg nilmanifold, associated with the compactification torus with a nontrivial flux condition on it, and the Riemann surface of genus one, related to the spatial worldvolume. 

Identifying low-dimensional nilmanifolds with torsion cycles in the M2-brane bundle description, is definitely, a way to characterize the torsion in the global description. Torsion cycles of order greater than one, also contribute to the equivalence classes of twisted torus bundles. From \cite{mpgm23} we know that the Hamiltonian of the QM2-brane on $\mathbb{R}\times T^2$ with parabolic monodromy is formulated on the coinvariants. This analysis was extended to elliptic and hyperbolic monodromies in \cite{mpgm24}. Nevertheless, the parabolic coinvariants considered in those works, are associated with parabolic monodromies \eqref{parabolicrep} with $k=1$. These monodromies correspond to torsion of order 1 in the homology. In this work, we show that the mass operator of the QM2-brane with parabolic monodromies, for all values of $k\in\mathbb{Z}$ is invariant on the coinvariants. The parabolic coinvariants for $k=1$ are divided into $k$ inequivalent coinvariants. The transformation within coinvariants, which is a symmetry of the mass operator, is given by a parabolic matrix of $\mbox{SL}(2,\mathbb{R})$. The transformation between inequivalent coinvariants is given by an $\mbox{SL}(2,\mathbb{R})$ matrix that generalizes the transformation considered in \cite{mpgm23,mpgm24}. The transformations considered in the aforementioned works are directly obtained by considering $k=1$, in the general transformations.

We also extend the analysis of \cite{mpgm23,mpgm24} to the coinvariant of the base manifold. These coinvariants are associated with an induced monodromy in the base. They can be thought of as equivalence classes of wrapping charges, which are understood as elements of $\mbox{H}^1(\Sigma)$. At the same time, they are related to the wrapping matrix, whose determinant is nonvanishing due to the flux condition, formerly known as central charge condition, that ensures the good quantum behavior of the theory. Consequently, for a given flux $n$, and a given $k$, the inequivalent coinvariants are finite and bounded by $n$. Indeed, the values of $q$ are restricted to the divisor of $n$. Nevertheless, by considering all possible values for $n$ and $k$, we consider all possible values of $q$. 

As the QM2-brane with parabolic monodromies is well-defined on the coinvariants, it makes sense to consider the double-dimensional reduction. We know that the case with $k=1$ leads to what we call $q$-strings, where $q\in\mathbb{Z}$ identifies the inequivalent coinvariants from the M2-brane perspective. Nevertheless, this case is associated with $k=1$. We found that for $k>1$, each coinvariant is split in $k$ coinvariants, and by considering the coinvariants of the base, only some coinvariants results consistent with the flux condition. Consequently, is natural to expect that, associated with each equivalence class, there is a state of parabolic ($p,q$)-string, whose low-energy limit corresponds to the parabolic gauge supergravity in nine dimensions.

\acknowledgments
MPGM is grateful to the Physics Dept. of Univ. Antofagasta (Chile) for their kind invitation, 
where part of this work was carried out. MPGM is partially supported by 
PID2021-125700NB-C21 MCI grant (Spain). CLH is partially supported by ANID 
posdoctorado Chile/2022-74220044. CLH is also grateful for the support of Project 
PID2021-123017NB-I00, funded by MCIN/AEI/10.13039/501100011033. AR thanks to SEM 18-02 project from U. Antofagasta.

\appendix
\section{Semi-direct products}
\label{Ape1}

There are two equivalent ways to describe semi-direct product constructions
\begin{enumerate}
    \item Let $G$ be a group with two subgroups $N$ and $H$ such that $N$ is normal. If $N\cap H=\left\{ e \right\}\subset G$ and every element of $G$ can be written as a product of an element of $N$ and an element of $H$, then $G=NH$. $G$ is a semi-direct product of its subgroups.
    \item Let $N$ and $H$ be two groups together with a left action oh $H$ on $N$, $\varphi=H\rightarrow \mbox{Aut}(N)$ which we denote by $\varphi_H(n)= ^hn$ for $h\in H$ and $n\in N$. The semi-direct product of $N$ and $H$ is the group $N\rtimes_\varphi H$ defined to be the set $N\times H$ with the product
    $$(n,h)(n',h')=(n \, {}^hn',hh').$$
    The inverse is then $$(n,h)^{-1}=(^{h^{-1}}n^{-1},h^{-1}).$$
\end{enumerate}
These two definitions are equivalent: Given $N,H\subset G$ as in point (1), then $G\approx N\rtimes_{\mbox{Ad}}H$ where $\mbox{Ad}_h(n))=hnh^{-1}$. On the other hand, every element of the group $G=N\rtimes_{\varphi} H$ defined in (2) can be written as $(n,h)=(n,e_H)(e_N,h)$ and the subgroups $N\times \left\{e_H\right\}$ and $\left\{e_N\right\}\times H$ intersect only in the identity of $G$. 

The relation between semi-direct product and quotient groups is given by the Lemma 2.3 from \cite{Aschieri}
\begin{lemma}
    Let $G=N\rtimes_\varphi H$ be a semi-direct product, and let $V\subset N$ be a subgroup which is normal in $G$. Then the quotient group $G/V$ is the semi-direct product $(N/V)\rtimes_{\varphi^V} H$, where $\varphi^V$ is the action of $\varphi$ of $H$ induced on the quotient group $N/V$.
\end{lemma}

% \paragraph{Note added.} This is also a good position for notes added
% after the paper has been written.


% \begin{thebibliography}{99}

% \bibitem{a}
% Author, \emph{Title}, \emph{J. Abbrev.} {\bf vol} (year) pg.

% \bibitem{b}
% Author, \emph{Title},
% arxiv:1234.5678.

% \bibitem{c}
% Author, \emph{Title},
% Publisher (year).

% % Please avoid comments such as "For a review'', "For some examples",
% % "and references therein" or move them in the text. In general,
% % please leave only references in the bibliography and move all
% % accessory text in footnotes.

% % Also, please have only one work for each \bibitem.

 % \bibliographystyle{unsrt}  
 %  \bibliography{ref}

\end{document}